%% file: main.tex
\documentclass[journal]{IEEEtran}
\usepackage{cite}
\usepackage{verbatim}
\usepackage{url}
\usepackage{amsmath}
\usepackage{url}
\usepackage{amssymb,amsbsy,epsfig,float,bm}
\usepackage{url}
\usepackage{amsmath}
\usepackage{graphicx}
\usepackage{multirow}
\usepackage{caption}
\usepackage{lipsum}
\usepackage{wrapfig}

\usepackage[caption=false,font=footnotesize]{subfig}

\newtheorem{thm}{Theorem}
\newtheorem{lemma}{Lemma}
\newtheorem{proposition}{Proposition}
\newtheorem{cor}{Corollary}
\newtheorem{remark}{Remark}
\newtheorem{proof}{Proof}

\newcommand{\G}{\gamma}
\newcommand{\OL}{\overline}
\newcommand{\LA}{\lambda}
\newcommand{\A}{\alpha}

\begin{document}

\title{Zero-rating of Content and its Effect on the Quality of Service in the Internet}

\author{Manjesh K.~Hanawal, Fehmina Malik
	and Yezekael Hayel
	\IEEEcompsocitemizethanks{\IEEEcompsocthanksitem Manjesh K. Hanawal and Fehmina Malik are with IEOR, IIT Bombay, India. E-mail: \{mhanawal, fmalik\}@iitb.ac.in. 	
		
		Yezekael Hayel is with LIA/CERI, University of Avignon, France. E-mail: yezekael.hayel@univ-avignon.fr 
	}
}

%\author{\IEEEauthorblockN{ Manjesh K. Hanawal, Fehmina Malik and  Yezekael Hayel}\\
%\IEEEauthorblockA{IEOR, IIT Bombay 
%Mumbai, India \\
%LIA/CERI, University of Avignon, France\\ 
%\{mhanawal, fmalik\}@iitb.ac.in, yezekael.hayel@univ-avignon.fr}
%}

\maketitle

\begin{abstract}
The ongoing net neutrality debate has generated a lot of heated discussions on whether
or not monetary interactions should be regulated between content and access providers. Among the several topics discussed, `differential pricing' has recently received attention due to `zero-rating' platforms proposed by some service providers. In the differential pricing scheme, Internet Service Providers (ISPs) can exempt data access charges for on content from certain CPs (zero-rated) while no exemption is on content from other CPs. This allows the possibility for Content Providers (CPs) to make `sponsorship'
agreements to zero-rate their content and attract more user traffic. In this paper, we study the effect of differential pricing on various players in the Internet. We first consider a model with a monopolistic ISP and multiple CPs where users select CPs based on the quality of service (QoS) and data access charges.  We show that in a differential pricing regime 1) a CP offering low QoS can make have higher surplus than a CP offering better QoS through sponsorships. 2) Overall QoS (mean delay) for end users can degrade under differential pricing schemes. In the oligopolistic market with multiple ISPs, users tend to select the ISP with lowest ISP resulting in same type of conclusions as in the monopolistic market. We then study how differential pricing effects the revenue of ISPs.
\end{abstract}

\section{Introduction}
\label{sec:Intro}
\input{Intro}
\section{Model and Setup}
\label{sec:Model}
\input{Model}

\section{User behavior}
\label{sec:UserBehavior}
\input{UserBehavior}

\section{Mean Delay in the Network}	
\label{sec:MeanDelay}
 \input{MeanDelay}
\section{Behavior of CPs}
\label{sec:CPBehavior}
\input{CPBehavior}
\section{Exogenous Arrival of Demand}
\label{sec:Exogenous}
\input{ExogenousArrivals}
\section{Non-cooperative Game with Actions \{S, N\}}
\label{sec:CPGame}
\input{CPGame}
\section{Revenue Gain for ISP}
\label{sec:RGF}
\input{RGF}
\section{Multiple ISPs}
\label{sec:MultiISP}
\input{MultiISP}
\section{Conclusions and Policy Recommendations}
\label{sec:Conclusions}
\input{Conclusions}

%\section*{Acknowledgment}
%
%This work is supported by funding from IIT Bombay IRCC SEED grant (16IRCCSG010), INSPIRE faculty fellowship (IFA-14/ENG-73) from DST, Govt. of India and Collaborative Scientific Research Programme by CEFIPRA (5702-1).

\section*{\center Appendices}
%\appendix
\input{Appendix}

\begin{IEEEbiographynophoto}{Manjesh K. Hanawal}
	received the M.S. degree in ECE from the Indian Institute of Science, Bangalore, India, in 2009,
	and the Ph.D. degree from INRIA, Sophia Antipolis, France, and the University of Avignon, Avignon, France, in 2013. After spending two
	years as a postdoctoral associate at Boston University, he is now an Assistant Professor in Industrial Engineering and Operations Research
	at the Indian Institute of Technology Bombay, Mumbai, India. His research interests include communication networks, machine learning
	and network economics.
\end{IEEEbiographynophoto} 

\begin{IEEEbiographynophoto}{Fehmina Malik}
 is currently pursuing Ph.D. at IEOR, IIT Bombay, Mumbai, India. She received her B.Sc. Hons. Mathematics degree,  M.Sc. and M.Phil in Operations Research from University of Delhi, Delhi, India in 2011, 2013 and 2015 respectively.  Her current research interests include Game theory, Internet Economics, Supply Chain and Inventory Management.
\end{IEEEbiographynophoto} 

\begin{IEEEbiographynophoto}{Yezekael Hayel}
%[{\includegraphics[width=1in,height=1.25in,clip,keepaspectratio]{yeze.eps}}]{Yezekael Hayel}(M'08) received the M.Sc. degree in computer science and applied mathematics from the University of Rennes 1 in 2002, and the Ph.D. degree in computer science from the University of Rennes 1 and INRIA in 2005.
	He has been Assistant/Associate Professor with the University of Avignon, France, since 2006. He has held a tenure position (HDR) since 2013. His research interests include performance evaluation and optimization of networks based on game theoretic and queuing models. He looks at applications in communication/transportation and social networks, such as wireless flexible networks, bio-inspired and self-organizing networks, and economic models of the Internet and yield management. Since joining the Networking Group of the LIA/CERI, he has participated in several projects. He was also involved in workshops and conference organizations. He participates in several national (ANR) and international projects with industrial companies, such as Orange Labs, Alcatel-Lucent, and IBM, and academic partners, such as Supelec, CNRS, and UCLA. He has been invited to give seminal talks in institutions, such as CRAN, INRIA, Supelec, UAM (Mexico), ALU (Shanghai), TU Delft, UGent and Boston University. He was a visiting professor at NYU Polytechnic School of Engineering in 2014/2015.  He is now the head of the computer science/engineering institute (CERI) of the University of Avignon.
\end{IEEEbiographynophoto}
\end{document}

%% file: Intro.tex
The term 'network neutrality'  generally refers to the principle that Internet Service Providers (ISPs) must treat all Internet traffic of a particular class on an equal basis, without regard to the type, origin, or destination of the content or the means of its transmission. What it implies is that all points in a network should be treated equally without any discrimination on speed, quality or price. Some of those features have made internet to grow at such a fast rate.  Any practice of blocking, throttling, preferential treatment, discriminatory tariffs  of content or applications is treated as non-neutral behavior. In this paper we study non-neutral behavior related to discriminatory access prices -- the price ISP charges the end users to provide access to the CPs.

In recent years, with the growing popularity of data intensive services, e.g., online video streaming and cloud-based applications, Internet traffic has been growing more than 50\% per year \cite{Sigcom10_InternetTraffic}, causing serious network congestions, especially in wireless networks which are bandwidth constrained. To sustain such rapid traffic growth and enhance user experiences, ISPs need to upgrade their network infrastructures and expand capacities. However, the revenues from end-users are often not enough to recoup the corresponding costs and ISPs are looking at other methods for revenue generation. Some methods being adopted are moving away from flat rate pricing to volume based pricing \cite{Wang17}, especially by the wireless ISPs. On the other hand, CPs that work on different revenue models (mainly advertisements driven) are seeing growth in revenues \cite{KearneyPaper} due to higher traffic. Pointing to this disparity, some ISPs have proposed that CPs return the benefit of value chain by sharing their revenues with them. In \cite{MishraViewPoint}, the author describes evolution of the pricing structure on the Internet along the different relationships between CPs, ISPs and end users. It is observed that:``the network neutrality issue is really about economics rather than freedom or promoting/stifling innovation."

While ISPs earnings depend on the total volume of traffic that flows through their networks (under volume based pricing), CPs revenue often depends on what fraction of that is directed to  it.  Also, as noted in \cite{MishraViewPoint}, there is a lot of competition on the content side of the Internet, but not enough at the last mile ISPs as most of times it is a monopolistic market. To increase their share of traffic, many CPs are exploring various smart data pricing (SDP) schemes \cite{CS2013_SDP} and also exploring favors from ISPs so that they stand out in the competition. Some CPs prefer that ISPs  incentivise the end users to access their content more either by giving higher priority to its content or exempting access charges on it. In turn, CPs can share the gains with the ISPs. This looks like an attractive proposition for ISPs who anyway wants CPs to share their revenue with them. Among the various models for monetary interactions, zero-rating has found traction. 

In zero-rating, an ISP and CPs enter into sponsorship agreement such that the ISP subsidizes data traffic charges (or access price) applicable on the content accessed from the CPs. The CPs compensate the ISP either by repaying the subsidy amount or offer free advertisement services. In return, sponsoring CPs hope to get more traffic and earn higher advertisement revenues. The article \cite{Ma16} describes an economic mechanism based on subsidization of usage-based access cost of users by the CPs. The author argue that this induces an improvement of the revenue of the ISP and thus strengthen the investment incentives.  Zero-rating of content has no effect on the end users under flat-rate pricing models. However it affects their decisions under volume based subscription where they are charged based on the amount of traffic they consume. We thus focus on volume based pricing models which is predominantly followed (or being planned) by the wireless ISPs. 
%The author first analyzes an economic model with one ISP, several CPs and end-users with no subsidizing mechanism. The different economic relationships between these actors are studied, and several properties of equilibrium situation in terms of utilization of the network are proposed. Second, the author introduces the subsidizing mechanism for CP and shows that it implies an investment incentives for the ISP because his revenue is improving.

Differential pricing schemes, and in particular zero-rating methods,  are non-neutral as it allows ISPs to discriminate content based on its origin. Net neutrality advocates argue that differential pricing hinders innovations at the CPs as new entrants cannot afford to get their content zero-rated and will be left out in the competition. Those favoring differential pricing argue that it helps more users connect to the Internet, especially in the developing countries where Internet penetration is still low. Some of zero-rated platforms like BingeOn by T-Mobile, FreeBee by Verizon, and Free Basics by Facebook are accused of violating net neutrality principles and are under scrutiny. The issue of differential pricing is now a part of consultations launched by several regulatory authorities including the FCC, European Commission, CRTC  \cite{EUConsultation, CanadaConsultation} seeking public and stake holders' opinion. Differential pricing is banned in Chile and The Netherlands with India being the latest to do so \cite{CRCOM, TRAI}.   

Zero-rating is a SDP scheme through which CPs aim to attract more user traffic. However, pricing schemes alone do not guarantee higher traffic as quality of service (QoS) also matters for users. While QoS experienced at CPs depends on long term planning and investment in the service capacities of their facilities, pricing strategies can be based on short term planning and constitute running costs. The CPs can tradeoff  between long term and running costs to maximize their revenues. Our aim in this paper is to understand how zero-rating schemes affect revenues of CPs and QoS experienced by the users. Specifically, we ask if a CP offering lower QoS can earn more revenue than CPs offering higher QoS through differential pricing? And, can QoS experienced by users under differential schemes degrade compared to the case where it is not allowed (neutral regime). 

As no market data is available for an empirical study of the effect of differential pricing, we take an analytical approach and model the scenario as follows: we consider a single ISP connecting users to a set of CPs that offer same/similar content. Users' decision to select a CP depends on the QoS and access price for the content at that CP. The users constitute mobile devices or any internet enabled devices that generate requests to access content from the CPs. The requests are  assumed to be generated according to a Poisson process. We consider a hierarchical game where ISP act as a leader and sets access price. The CPs then negotiate with the ISP and competitively decide what fraction of access prices they will sponsor (or subsidize) such that their utility, defined as the difference between the average revenues earned from the user traffic and the amount they have to pay to the ISP, is maximized. Finally, knowing CPs decisions users compete and select CPs such that their cost, defined as the sum of mean delay and the corresponding access price, is minimized. We analyze the hierarchical game via a backward induction technique. 

Our analysis reveals that answers to both previous questions can be positive. We identify the regimes where differential pricing schemes lead to unfair distribution of revenues among CPs and users QoS experience degrades. However, if access price set by ISPs are regulated, both the unfavorable scenarios can be avoided.

The rest of the paper is organized as follows: In Section \ref{sec:Model}, we discuss the model and settings of the hierarchical game. We begin with equilibrium behavior of the users in Section \ref{sec:UserBehavior} and study mean delay experienced by them in Section \ref{sec:MeanDelay}. In section \ref{sec:CPBehavior}, we study the preference of CPs for the differential pricing and demonstrate that the game need not have Nash equilibria. In Section \ref{sec:Exogenous}, we consider the exogenous demand and study its effect on the CPs' behavior. In Section \ref{sec:CPGame}, we analyze the game where CPs decisions are restricted to either fully sponsor the access price  or do not sponsor at all. We study the resulting monetary gain for the ISP at equilibrium in Section \ref{sec:RGF}. We extend our analysis to include multiple ISPs in Section \ref{sec:MultiISP}. Finally, we discuss regulatory implications of our analysis and future extensions in Section \ref{sec:Conclusions}.

\subsection{Literature review}
There is a significant amount of literature that analyzes various aspects of the net neutrality debate, like incentive for investment, QoS differentiation, side payments or off-network pricing through analytical models. For a detailed survey see \cite{Altman11}. However, the literature on the effect of differential pricing and QoS experienced by users in this regime are few and we discuss them below.

In \cite{Infocom13_SponsoringContent}, the authors study a game between a CP and an ISP where the ISP first sets the price parameters for sponsorship and the CP responds by deciding what volume of content it will sponsor. This model assumes that the users always access the sponsored content irrespective of the QoS which is not always true. Multiple CPs involving larger (richer) and smaller revenues are considered in \cite{Infocom14_SponsoringContent}. When the ISP charges both the end users and the CPs, it is argued that richer CPs  derive more benefit through sponsorship in the long run (market shares are dynamic). Negotiation between the ISP and CPs are studied in \cite{La1} using Stakelberg game or Nash bargaining game where CPs negotiate with the ISP for higher QoS for its content. It is argued that QoS at equilibrium improves in both the games.  The analysis involving stakelberg game is extended to multiple CPs in \cite{La2}. A scheme named Quality Sponsor Data (QSD) is proposed  in \cite{WiOpt15_SponsoringContent} where ISPs make portion of their resources available to CPs for sponsorship. Voluntary subsidization of the data usage charges by CPs for accessing their content is proposed in \cite{CoNEXT_Ma}. It is argued that voluntary subsidization of traffic increases welfare of the Internet content market. Hierarchal games involving ISP, CPs and multiple type of users is analyzed in \cite{Infocom15_SponsoringData} and show that all parties benefit form sponsored data. However, the work ignores the effect of QoS.  In \cite{Sigmetrics15_SponsoredData}, QoS parameter are considered in deriving the total traffic generated by the users. A stakelberg game between an ISP and CPs are analyzed where the ISP decides the access price and the CPs decide whether or not to sponsor user traffic. Competition among the users and their strategic behavior is not considered in this work. Furthermore, this work considers `QoS index' as an exogenous parameter (required bandwidth), whereas we consider perceived QoS by the users. In \cite{Zou17}, the authors study ISPÕs optimal service differentiation strategy subject to its network capacity constraints. Based on optimal taxation theory, the authors consider an optimal control problem. In \cite{Jullien18}, the authors propose a model considering sponsored data plans in which content providers pay for traffic on behalf of their consumers. In this paper, the authors consider a demand-response type model for end users consumption whereas, in our framework we assume a congestion game in order to model the interaction through a quality metric, that impacts directly the demand. Agreements between CPs about linking their contents and there propose a better offer is analyzed in \cite{calzada18} through the network neutrality prism. The authors show that CPs are interested in reaching a linking agreement when the termination fee set by the Internet Service Provider (ISP) is not particularly high. In their model, the authors do not also consider explicit congestion at the demand side, as we consider in our analysis. Moreover, we do not consider agreements between CPs about content, we assumer only possible economic agreements between CPs and one ISP.

Our work differs from all the models studied in the literature as we consider users to be strategic and  traffic distribution in the network  is derived from their equilibrium behavior. Further, we focus  on QoS offered at the CPs and study how it influences the CPs sponsorship decisions and how it in turn affects the revenue of the ISPs. This work is an significant extension of an earlier conference version \cite{WiOpt18} that only considered the single ISP case with preliminary discussions about exogenous arrivals. 

%% file: Model.tex
We first consider a monopolistic market with a single Internet Service Provider (ISP) that connects end users to the Internet. We focus on a particular type of non-elastic traffic like content, say videos, music, or online shopping, that the users can access from content providers (CPs). Multiple CPs offer same/similar type of contents and end users can access content from any of them. The ISP charges $c$ monetary units per unit of traffic accessed through its network to the end users. We refer to this price as the `access price'. As specified in \cite{Wang17}, recently, broadband ISPs in US and Europe introduce a data-cap and adopt a two-part tariff structure, a combination of the flat-rate and usage-based pricing. Under such a two-part scheme, additional charges are imposed if a user�s data usage exceeds the data cap and the exceeded amount is charged based on a per-unit fee. Therefore, we consider in our model the usage-based part of the scheme which integrates the relationship between access price and demand. Let $N$ denote the number of CPs and $[N]$ the set of CPs. Each CP can enter into zero-rating-agreement with the ISP. In this case, CP pays a proportion of the access price for the content accessed from it and the ISP passes on the benefit to the end users. This proportion may be different for each CP. Let $\gamma_i \in [0\; 1],i \in[N]$ denotes the fraction of the access price end users pay to access content of $i$-th CP (also denoted $CP_i$). The value of $\G_i, i\in[N]$ is decided by $i$-th CP and we refer to it as the {\it subsidy factor}. For every unit of traffic accessed from $i$-th CP, end users and $i$-th CP pay $\gamma_ic$ and  $(1-\gamma_i)c$, respectively, to the ISP.  We allow the possibility for $i$-th CP to pay the ISP only $\rho \gamma_ic$ per unit traffic accessed from it, where $\rho \in (0\;1 ]$. This parameter determines the level of negotiation between CPs and the ISP. We refer to the special cases when $\gamma_i \in (0\; 1)$, $\gamma_i=0$, and $\gamma_i=1$, as $i$-th CP is Partially-sponsoring (P), Sponsoring (S), and No-sponsoring (N), respectively.

\begin{figure}
	\centering
	\includegraphics[scale=.36]{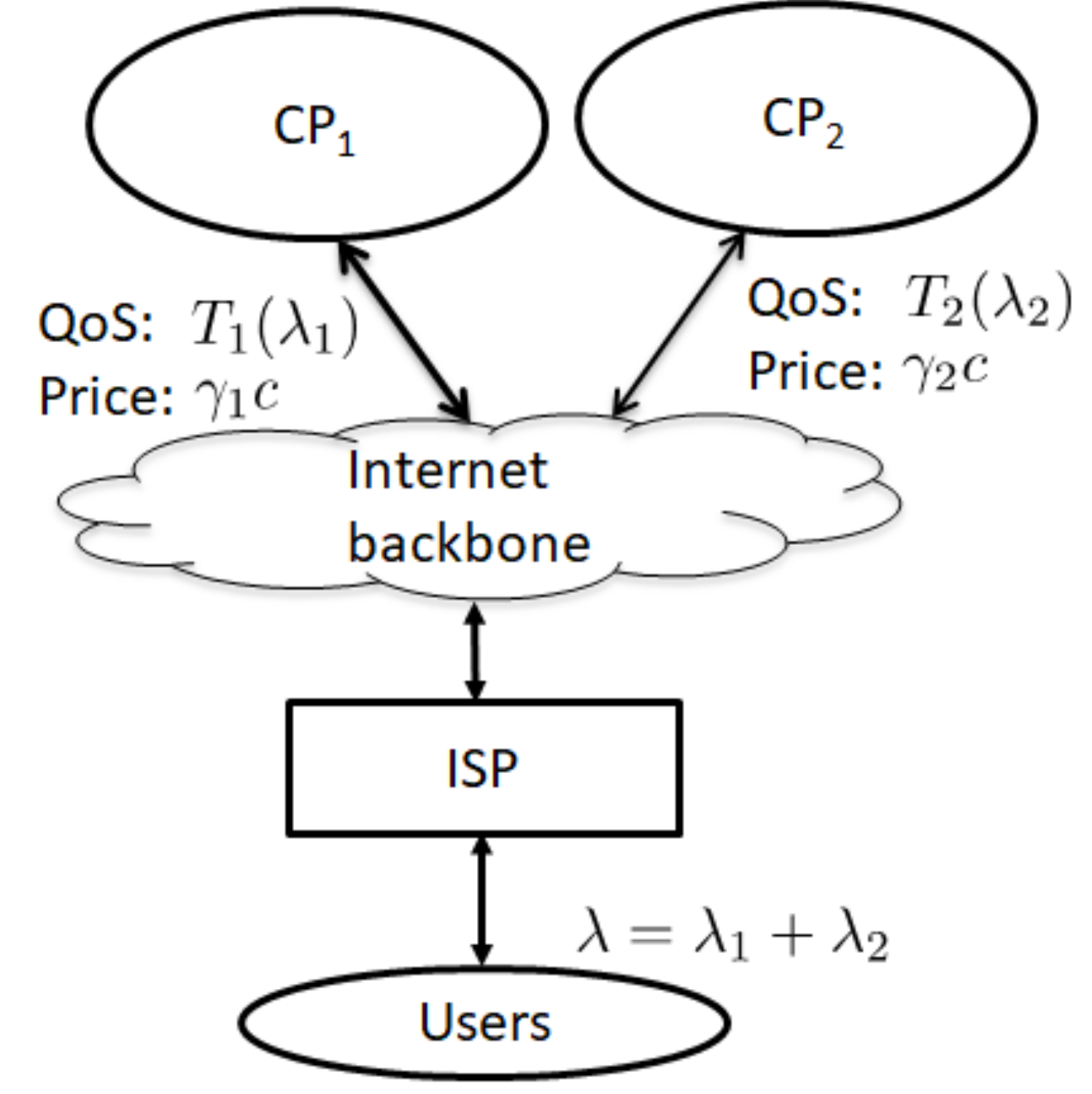}
	\caption{ Interaction between the ISP, CPs and end users.}
		\label{fig:Connections}
\end{figure}

We consider a large population of end users that access the content from CPs by sending requests to them. The requests are generated according to Poisson process with a rate $\lambda$. Each request results in a certain amount of traffic flow between the CPs and end users going through the ISP. Without loss of generality, we assume that the mean traffic flow per request is one unit\footnote{Otherwise $\rho$ can be appropriately scaled. See (\ref{eqn:CPGUtility}).}. Each end user decides which CP to send its request and this selection does not depend on the choice of other users and its past decisions\footnote{The assumption that users do not have memory is not overly restrictive as the population is large and each new request is likely from a different user.}. The quality of the service (QoS) experienced by end users depends on the quantity of requests served and the service capacity of the CPs. The later depends on the investments in infrastructure. Higher the investments more is the capacity. Let $m_i, i\in [N]$ denotes the service capacity of $i$-th CP and let $T_i(\lambda_i)$ denotes the QoS experienced by end users at $i$-th CP when the CP is serving $\lambda_i$ requests. We stress that the QoS here refers to that experienced at the CPs and not on the Internet backbone or the ISP network. The network model for the case of two CPs is depicted in Figure \ref{fig:Connections}.

\subsection{Utility of Users}
Each end user likes its request to be served as early as possible. We consider the mean delay experienced at each CP as QoS metric and set $T_i(x)=\frac{1}{m_i -x}$ for all $i\in [N]$, which is a well known mean (stationary) delay value in a $M/M/1$ queue \cite{Kleinrock}\footnote{Note that if the service times at the CPs are exponentially distributed, then $1/(m_i-x)$ gives the exact value of mean delay at $i$-th CP when it receives request at Poisson rate $x$.}.  End users' decision on which CP to send their request depends on the QoS as well as the access price. We define utility (or cost ) of each end user by selecting a CP as the sum of mean delay and access price it incurs at that CP. Specifically, the cost of an end user served at $i$-th CP with requests at rate $x$ is\footnote{A more general cost function $C_i(x)$ that is convex and strictly increasing in $x$ can also be analyzed using techniques we use in this paper.}
\begin{equation}
\label{eqn:UserUtility}
C_i(x):=C_i(x,m_i,c,\gamma_i)=\begin{cases}
\frac{1}{m_i -x}  + \gamma_ic &\mbox{  if } m_i > x, \\
\infty &\mbox{otherwise.}
\end{cases}
\end{equation}
Each end user aims to select a CP that gives the minimum cost. Note that this formulation is the same as allowing each user to choose a CP with the lowest mean delay subject to a price constraint. This criteria also turns out to be optimizing weighted sum of mean delay and price with the weight corresponding to the Lagrangian multiplier associated to the constraint.

\subsection{Utility of Content Providers}
The revenue of CPs is an increasing function of the number of requests they receive. Generally, CPs revenue comes from per-click advertisements accessible on their webpages and higher requests imply more views of advertisements (for example, CP can embed an advertisement in each request served) resulting in higher income. We define utility of $CP_i$ as the net revenue earned after discounting the access price through sponsorship. Let $U_i$ denote the utility of $CP_i$, $ i\in [N]$, when it sponsors $\gamma_i$ fraction of access price and receives $\lambda_i$ mean number of requests, then: 
\begin{equation}
\label{eqn:CPGUtility}
U_i(\G_i)= f_i(\lambda_i) -\rho (1-\gamma_i)c\lambda_i.
\end{equation}
The function $f_i(\cdot)$ is monotonically increasing and denotes the revenue from per-click advertisements for example. The definition of this function is out of the scope of this paper. The number of requests received at a CP is the function of the $\gamma_i, i\in [N]$ set by all the CPs. End users react to these decisions, and hence, the decision of CPs are coupled through end users behavior. We assume that the value of $m_i, i\in [N]$ are fixed and are public knowledge.

\subsection{Decision Timescale and Hierarchy}
We now explain decision stages of our complex system which involves different types of players:  ISP, CPs and end users. Each of them makes decisions in hierarchy and their decisions change over different timescales. At the top of the hierarchy the ISP sets the access price $c$ which remains fixed for a long duration. Next in the hierarchy CPs make sponsorship decisions after knowing the access price set by the ISP. The CPs decision (parameter $\gamma_i$) can change at a timescale that is smaller than that of the ISP. At the bottom level of the hierarchy the end users decide from which CP to access content knowing subsidized access prices for all CPs. The users' decisions change at a smaller timescale than that of the CPs.

In the following we study the game among each type of players and how decisions of each type influences the performance of the other players. We make the following assumptions in the rest of the paper:
\begin{enumerate}
	\item[A1:] $\sum_{i=1}^{N} m_i > \lambda$.
	\item[A2:]  $m_i < \lambda$ for all $i\in [N]$.
	\item[A3:]  $m_1 < m_2 \leq m_3 \ldots \leq m_N$ and $m_1 > \lambda/N$.
\end{enumerate}
The first assumption ensures that a user can always find a CP where mean waiting time is bounded. The second assumption implies that no single CP is capable of handling all the requests alone and hence requests get spread across multiple CPs. The last condition indexes the CPs according to their rank in service capacity and ensures that the CP with lowest service capacity gets non-zero amount of requests at equilibrium. For notational convenience, we write $\overline{m}=\sum_{i=1}^N m_i - \lambda$, which denotes the excess service capacity in the network.

To maintain analytic tractability and get clear insights, we restrict our theoretical analysis to the case with two CPs, i.e., set $N=2$. However, for the general $N$ numerical studies can be done as it turns out that equilibrium rates can be found generally efficiently solving a convex program (see (\cite{FrankWolf})). We start with the study of end users behavior.

%% file: UserBehavior.tex
In this section, we study how end users respond to sponsorship decisions by the CPs. Each user selects one CP to process its request without knowing current occupancy of the CPs nor the past (and future) arrival of requests from other end users. Then, the end users' decision to select a CP is necessarily probabilistic and users aim to select CPs according to a distribution that results in smallest expected cost.  Since all users are identical, we are interested in symmetric decisions where every user applies the same probabilistic decision on arrival of a new request. For any symmetric decision, arrival rate of requests at each CP also follows a Poisson process by the thinning property of the Poisson processes.  
%A probabilistic selection of CPs is said to be a symmetric Nash equilibrium if no end user has an incentive to deviate from it unilaterally. 
%Let $\lambda^*=(\lambda_1^*, \lambda_2^*, \ldots,\lambda_N^*)$, where $\lambda_i^*:=\lambda_i^*(\G_1,\G_2,\ldots, \G_N), i\in[N]$, is the arrival rate at $i$-th CP and $\sum_{i=1}^N\lambda_i^*=\lambda$, denote a symmetric Nash equilibrium of the game. 

We apply Wardrop equilibrium conditions to find the equilibrium decision strategy of large population of users\cite{Wardrop}. Let $\gamma_{-i}$ denote the subsidy factor of the CP other than $CP_i$ and  $\lambda^*=(\lambda_1^*, \lambda_2^*)$, where $\lambda_i^*:=\lambda_i^*(\G_i,\G_{-i}), i\in[N]$, is the arrival rate at $i$-th CP at equilibrium. we have that $\sum_{i=1}^N\lambda_i^*=\lambda^*$. Then we have the following properties coming from Wardrop principles that can be summarized in our context to the following sentence: "Every strictly positive request implies minimum cost" \footnote{We can apply Wardrop equilibrium conditions as the request arrivals satisfy the PASTA (Poisson Arrivals See Time Averages) property \cite{PASTA}.}:
\begin{equation}
\forall \;\; \lambda_i^*>0 \;\;\implies \;\; C_i(\lambda_i^*)\leq  C_j(\lambda_j^*) \;\; \mbox{ for  all} \quad j\neq i,
\end{equation}
which can also be expressed as
\begin{equation}
\forall \; i, \;\; \lambda_i^*\cdot (C_i(\lambda_i^*) - \alpha)=0, \mbox{ where } \alpha= \min_{i} C_i(\lambda_i^*).
\end{equation}
%For any given $N$, $\lambda$ and $C_i(\cdot), i\in N$, the Wardrop equilibrium can be computed as the unique solution of the following convex optimization problem:
%\begin{align}
%\label{eqb:OptimN}
%\min_{(\lambda_1,\ldots\lambda_N)} & \sum_{i=1}^{N}\int_{0}^{\lambda_i} C_i(y) \rm {d}y \\
%\nonumber
%\mbox{subjected to} &\sum_{i=1}^{N} \lambda_i =\lambda \mbox{  and  } \lambda_i>0 \;\; \forall \; i.
%\end{align}
%Note that when functions $C_i(\cdot)$ are convex, the above program is convex and Wardrop equilibrium can be efficiently computed using convex programs like the Frank-Wolfe algorithm \cite{FrankWolf}. The following proposition gives explicit characterization of Wardrop equilibrium.

\begin{lemma}
\label{lma:EquRates}
For $N=2$ and a given action profile $(\G_1,\G_2)$ of the CPs,  the equilibrium rates are as follows:
\begin{equation}
\label{eqn:EquRates}
\lambda_i^* = m_i - \frac{1}{\alpha-\gamma_i c} \;\;\forall\;i=1,2,
\end{equation}
where $\alpha:=\alpha(\gamma_1,\gamma_2)$ is the equilibrium cost given by
\begin{equation}
\label{eqn:MinCost}
\alpha=\frac{c(\gamma_1+\gamma_2)}{2}+\frac{1}{\overline{m}}+ \sqrt{\frac{(c(\gamma_1-\gamma_2))^2}{4}+\frac{1}{\overline{m}^2}}.
\end{equation}
\end{lemma}

The following lemma follows immediately the previous lemma.
\begin{lemma}
\label{lma:EquRateMonotone}
For given $\G_{i}, i=1,2$, the equilibrium rate $\lambda_i^*$ is strictly increasing in $\G_{-i}$, i.e., equilibrium rates at a CP increases if the subsidy factor for other CP increases. %\hfill\IEEEQED
\end{lemma}

From equation (\ref{eqn:MinCost}), it is also clear that the equilibrium cost $\alpha$ is monotonically increasing in $c$. However, the effect of $c$ on the equilibrium rates depends on the values of $(\G_1,\G_2)$ relative to each others. We have the following proposition that describes monotonicity properties of the equilibrium rates depending on the subsidy factors set by the CPs. 

\begin{proposition}
\label{prop:EquRateProperty}
For any $(\G_1,\G_2)$, equilibrium rates satisfy the following properties:
\begin{enumerate}
	\item if $\G_1<\G_2$, then $\LA_1^*$ is monotonically increasing in $c$ (and $\LA_2^*$ is decreasing in $c$).
	\item if $\G_1\geq\G_2$, then $\LA_1^*$ is monotonically decreasing in $c$ (and $\LA_2^*$ is increasing in $c$).%\hfill\IEEEQED
\end{enumerate}
\end{proposition}

Interestingly, when the sponsorship decisions of the CPs are symmetric, i.e., all CPs set the same subsidy factor, users are indifferent to the sponsorship decisions. To see this, notice from (\ref{eqn:MinCost}) that when $\G_i=\G\;\; \forall i=1,2$  for some $\G \in [0\;1]$ the equilibrium rates in (\ref{eqn:EquRates}) are independent of $\G$. Thus, if the access prices for the content across all the CPs if either increased or decreased by the same amount, users preferences for the CPs do not change.

%% file: MeanDelay.tex
In this section, we analyze the average waiting time (or delay) experienced by a 'typical' end user at equilibrium situation. We refer to any user that arrives at equilibrium as a typical end user.

Given a global rate $\lambda$, a fraction $\lambda_i^*/\lambda$ of the users' requests are served at $i$-th CP at equilibrium, and each one of them incur mean delay of $1/(m_i-\lambda_i^*)$. Since the end users are homogeneous, the equilibrium strategy of each end user is equivalent to the strategy to select $CP_i$ with probability $\lambda_i^*/\lambda$. Hence we define mean delay (henceforth referred simply as delay) experienced by a typical user as: 
\begin{equation}
\label{eqn:Delay}
D(c,\gamma_1,\gamma_2)=\sum_{i=1}^{N}\frac{\lambda_i^*}{\lambda}\frac{1}{m_i- \lambda_i^*}.
\end{equation} 
\begin{lemma}
\label{lma:DelaySimplify}
Given $(\gamma_1,\gamma_2)$ and $c$, we have 
\begin{equation}
\label{eqn:DealySimpliefied}
D(c,\gamma_1, \gamma_2)=\frac{\alpha}{\lambda}\sum_{i=1}^2m_i - \frac{c}{\lambda} \sum_{i=1}^2m_i\gamma_i -\frac{2}{\lambda}, 
\end{equation}
where $\alpha$ is given by Equation (\ref{eqn:MinCost}). And when $\gamma_i=\gamma$ for all $i=1,2$, for some $\gamma \in [0\;1]$ we get 
\begin{equation}
\label{eqn:DealySymmetric}
D(c,\gamma):=D(c, \gamma, \gamma)=\frac{2}{\overline{m}}. 
\end{equation}
%\hfill\IEEEQED
\end{lemma}
When $\gamma_i$ are the same across all CPs, i.e., sponsorship decisions are symmetric, then the delay does not depend neither on the price set by the ISP nor on the subsidy factor set by the  CPs. In this case, delay only depends on the excess capacity-- larger the excess capacity smaller the delay. When $N>2$, it can be shown that the delay is given by $N/\overline{m}$, i.e., increases linearly with the number of CPs. 
\noindent
%When the sponsorship decisions are asymmetric, delay is a function of $c$.
\begin{lemma}
	\label{lma:DelayProp}
For each couple $(\gamma_1,\gamma_2)$, the delay  $D(c, \gamma_1,\gamma_2)$ is convex in  $c$. Further, if $\gamma_1 \geq \gamma_2$ it is monotonically increasing in $c$.
\end{lemma}

An heuristic argument is as follows. Recall the assumption that $m_1< m_2$. When $\G_1>\G_2$, the equilibrium arrival rate increases at $CP_2$ (see Prop. \ref{prop:EquRateProperty}) with $c$ which in turn increases delay experienced by a typical user at $CP_2$. However, the corresponding rate of delay decrease at $CP_1$ (from the decreased arrival rate) is smaller due to its smaller capacity and it results in a overall increase in delay. On the other hand, when $\G_1<\G_2$, the equilibrium arrival rate increases at $CP_1$ (see Prop. \ref{prop:EquRateProperty}) with $c$ which in turn increases delay for a typical user at $CP_1$ but decreases at $CP_2$. For smaller values of $c$, the rate of increase in delay at $CP_1$ is small compared to the rate of decrease in delay at $CP_2$ as $m_2>m_1$. However, as more arrivals shift to $CP_1$ with larger value of $c$, the delay increases significantly at $CP_1$ and can dominate the rate of decrease of delay at $CP_2$. Hence, with increasing value $c$, delay first decreases and then can increase. The threshold on $c$ where the delay changes its behavior, depends on the gap between service capacity of the CPs, i.e., $m_2-m_1$; larger the gap more prominent is the rate of decrease of delay at $CP_2$ and hence decrease in delay continues over a larger values of $c$. 

%The above lemma suggests that the users' preference for a CP depends on the difference between the access prices for their content, i.e., $(\G_1-\G_2)c$. If the difference gets larger, users' preference for one of the CP increases -- positive gaps implies preference for $CP_2$, and negative gap implies preference for $CP_1$.   

In the following, we study mean delay experienced by a typical user when the decisions of CPs are asymmetric and compare it with the case when decision of the  CPs are symmetric. Since mean delay is invariant to the amount of subsidy in the later case, we consider symmetric action profile $(N,N)$ (corresponding to $\gamma_1=\gamma_2=1$) as a reference and treat it as a regime with no-differential pricing or `neutral'. The main result of this section follows. 
\begin{thm}
	\label{thm:DelayProp}
For any $(\G_1,\G_2)$, the following properties hold:
\begin{enumerate}
	\item If $\G_2\leq \G_1$, then $D(c,\G_1,\G_2)\geq D(c, 1, 1)$ for all $c$.
	\item If $\G_2>\G_1$, then $D(c,\G_1,\G_2)\geq D(c, 1, 1)$ if and only if
	\begin{equation}
	\label{eqn:DelayThreshold}
     c(\G_2-\G_1) \geq \frac{1}{\overline{m}}\left (\frac{m_2}{m_1}-\frac{m_1}{m_2}\right ).
     \end{equation}
\end{enumerate}
\end{thm}
Notice that $D(0, \gamma_1, \gamma_2)=D(c,\gamma,\gamma)$ for all $\gamma_1, \gamma_2, \gamma \in [0\;1]$ and $c$. On one hand, for the case $\G_2 \leq \G_1$, it is shown in lemma \ref{lma:DelayProp} that  the delay is increasing in $c$, hence the first assertion in the theorem implying that delay in the differential regime is always higher compared to the neutral regime. On the other hand, when $\G_2 >\G_1$, delay experienced by a typical user is larger than in the neutral regime provided access price is larger than a certain threshold, otherwise it will be smaller. To understand this behavior notice that $D(c, \G_1, \G_2)$ is convex in $c$ (see Lemma (\ref{lma:DelayProp})). This indicates that there exists a threshold on $c$ above which delay will be higher than in the neutral regime. As given in (\ref{eqn:DelayThreshold}) this threshold increases if disparity between the service capacities of the CPs ($m_2/m_1$) increases and/or disparity between the access prices ($\G_2-\G_1$) for the content of the CPs decreases.

In summary, the above result suggests that the differential pricing can be unfavorable to the users. Specifically, when the CPs sponsorship decisions are asymmetric the mean delay for a typical user can be higher than that in the neutral regime. 

Theorem \ref{thm:DelayProp} also explains the effect of access price set by the ISP on delay experienced by the users. When $\G_2> \G_1$, smaller access prices actually benefits the users as delay in the differential regime is smaller compared with the neutral regime. However, if the access prices is large, more than a certain threshold, this favorable scenario no longer holds delay in the network can be higher. 
\begin{remark}
	In a differential regime, the QoS for end users can be degraded compared to that in the neutral regime if the CP with poor QoS offers higher subsidy than the CP with better QoS.
	\end{remark}

%% file: CPBehavior.tex
In this section, we analyze the behavior of CPs in this hierarchical system. Particularly,  we are interested in the question of which type of CPs prefer differential pricing or zero-rating scheme. We study particularly if a CP with lower service capacity can earn more revenue than the other CP by allowing higher subsidy. We will then study competition between the CPs and analyze their equilibrium behavior. 

In the following we assume that the advertisement revenue of each CP is proportional to their request arrival rate.  We set $f(\lambda_i)=\beta \lambda_i$, where $\beta >0$ is a constant that depends on how the traffic translates to revenues at the CPs which can be obtained from statistical analysis of per-click data usage. Utility of each CP depends on the vector $(\lambda_1^*,\ldots,\lambda_N^*)$ of equilibrium rates and it is defined by (with abuse of notation):
\begin{equation}
	\label{eqn:CPUtility}
	U_i(\G_i, \G_{-i})=(\beta- (1-\G_i)\rho c)\lambda_i^*.
\end{equation}
Clearly, $CP_i$ will sponsor its content if and only if $(1-\G_i)<\beta /  c\rho $.

\subsection{Which CPs Prefer Differential Pricing? }
In the differential pricing regime $CP_1$ can attract more user traffic than $CP_2$ by offering higher subsidy. However, higher subsidy may also increase the amount $CP_1$ has to pay to ISP. Then, a natural question is the following: can $CP_1$ set its subsidy factor such that it gets higher utility than $CP_2$? The following proposition gives the condition for this to happen.
\begin{proposition}
	\label{prop:CP1larger}
Let $\overline{\beta}=\rho c/\beta\leq 1$. For given $\G_1, \G_2$, we have that $U_1(\G_1,\G_2)  \geq U_2(\G_1,\G_2)$ if and only if
\begin{equation}
\label{eqn:U1Better}
\lambda_1^* \geq\frac{ \lambda}{\frac{1-(1-\G_1)\overline{\beta}}{1-(1-\G_2)\overline{\beta}}+1}.
\end{equation}
\end{proposition}
\begin{figure*}[!h]
	\centering
		\subfloat[]{\includegraphics[scale=0.23]{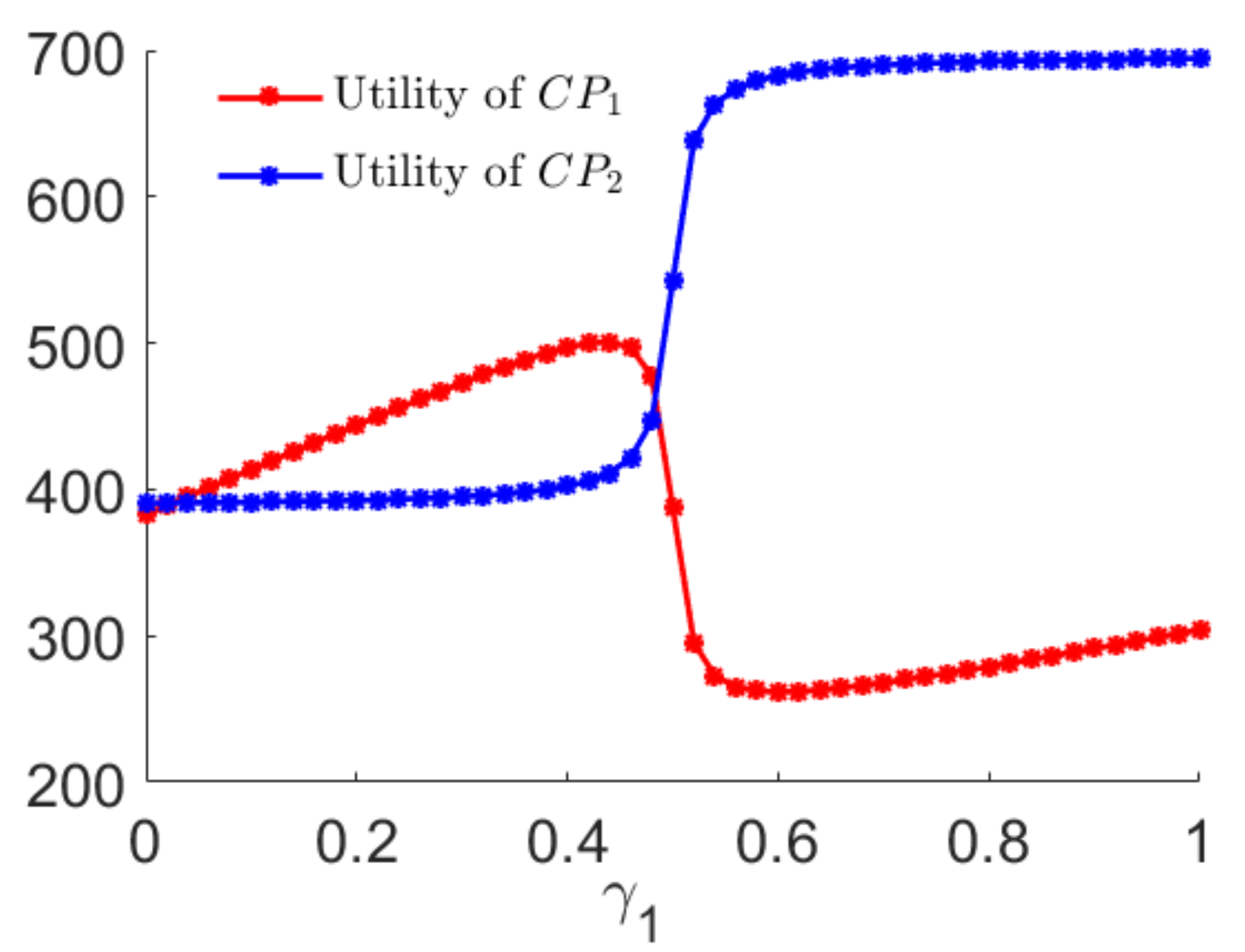}%
		\label{fig:Utility1}}
	\hspace{1cm}
	\subfloat[]{\includegraphics[scale=0.23]{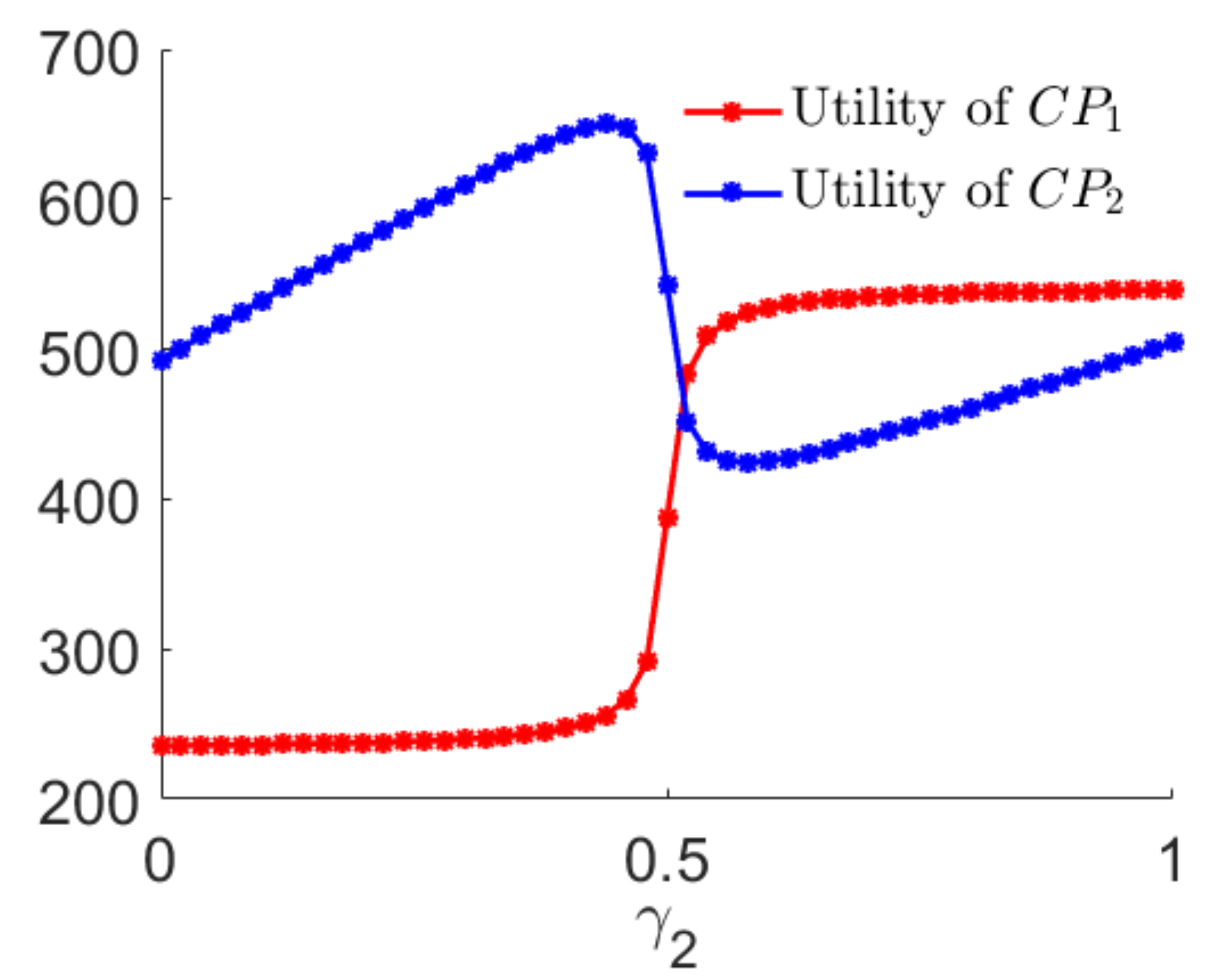}%
		\label{fig:Utility2}}
	\hspace{1cm}
	\subfloat[]{\includegraphics[scale=0.23]{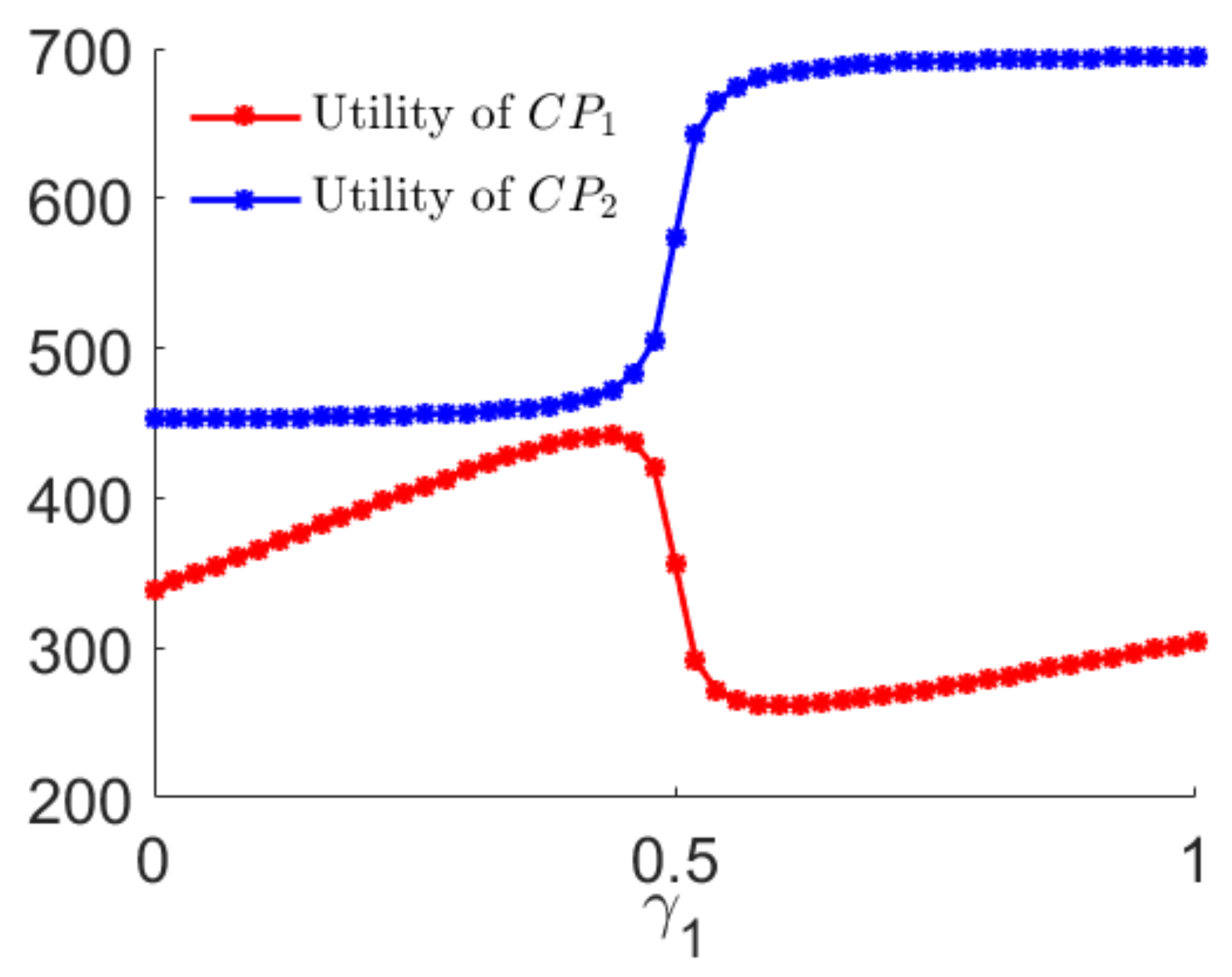}%
		\label{fig:Utility3}}
	\caption {The utilities of $CP_1$ and $CP_2$ are compared with parameters $\beta=1,\rho=.9, c=0.5, \lambda=1200$. In Fig. \ref{fig:Utility1} capacities of CPs are close, we set $M_1=700, M_2=900$ and $\gamma_2=0.5$. Here, utility of $CP_1$ is higher than that of $CP_2$ for small values of $\G_1$. In \ref{fig:Utility2}, we use the same parameters as in Fig. \ref{fig:Utility1} with $\gamma_1=0.5$. In Fig. \ref{fig:Utility3},  service capacity utility of $CP_1$ much lower than that of $CP_2$, we set $M_1=620, M_2=900$ and $\gamma_2=0.5$. Here, utility of $CP_1$ is always lower  than that of $CP_2$.}
	\label{fig:UtilityComp}
\end{figure*}
The technical condition $\bar{\beta}<1$ allows the CPs to consider all subsidy factors in the interval $[0,1]$. If this condition is not met, CPs have to keep subsidy factor over a restricted range to get positive utility. Note that $\lambda_1^*$ and the ratio on the right hand of the above condition are both decreasing in $\G_1$. The rate of decrease of $\lambda_1^*$ is lower when $\G_1$ is small and the above condition holds. Indeed, a numerical example in Fig. \ref{fig:Utility1} demonstrates that the condition holds. In Fig. \ref{fig:Utility3} we also show a case where the condition does not hold. In particular, when the difference between service capacities of the CPs is relatively small, condition (\ref{eqn:U1Better}) holds for some $\G_1$, and, in spite of its inferior service capacity $CP_1$ can earn higher utility. However, when the difference between service capacities is large, $CP_1$ cannot get higher utility than $CP_2$ through higher subsidy.

%
% \begin{figure}[!h]
% 	\begin{minipage}[t]{0.5\linewidth}
% 		\centering
% 		\includegraphics[scale=.2]{figures/CPUtilityComp}
% 		\caption{Utility of $CP_1$ vs $CP_2$}
% 		\label{fig:UtilityComp1}
% 	\end{minipage}
% 	\hspace{0.1cm}
% 	\begin{minipage}[t]{0.5\linewidth} 
% 		\centering
% 		\includegraphics[scale=.2]{figures/CPUtilityComp2}
% 		\caption{Utility of $CP_1$ vs $CP_2$}
% 		\label{fig:UtilityComp2}
% 	\end{minipage} 
% The figures demonstrate theat the $CP_1$ who has lower service capacity than $CP_2$ can earn more
% revenue by setting offering larger discounts on the access price.       
% \end{figure} 
%

These examples demonstrate that  the differential pricing regime can de-incentivizes the CPs from increasing their capacity resulting in degraded QoS for the end users. Specifically, the CPs with lower service capacity can prefer to offer higher subsidy rather than investing in their infrastructure and still end up earning higher revenue that CP that have invested more in their infrastructure but are not offering higher subsidy.

%\begin{remark}
%	$\rho c/\beta$ cannot be greater than 1, because it will make utility of CPs negative.
%\end{remark}
\begin{remark}
In differential pricing regime, a CP with lower QoS can earn more revenue than a CP with higher QoS by offering appropriate subsidy on access price.
\end{remark} 

\subsection{Non-Cooperative game between CPs}
The CPs aim to maximize their utility by appropriately setting their subsidy factor $\gamma_i$. We study the interaction as a non cooperative game between the CPs where the action of the $CP_i$ is to set a subsidy factor $\G_i$  that maximizes its utility. The CPs know that the end users select a CP to serve their request based on quality and the (subsidized) access price they offer. The objective of the $i$-th CP is given by the optimization problem:
$\max_{\G \in [0\;\;1]} U_i(\gamma, \G_{-i}) $.

\noindent
We say that $(\G_1^*, \G_2^*)$ is a Nash equilibrium for the non-cooperative game between the CPs if 
 $U_i(\G_i, \G^*_{-i}) \leq U_i(\G^*_i, \G^*_{-i})  \; \forall i$ and $\gamma_i \in [0\;1]$. For a given $\G_{-i}, i=1,2$, let $\overline{\G}_i(\G_{-i})$ denotes the best response of $CP_i$, i.e., 
 $\overline{\G}_i(\G_{-i})\in \arg\max_{\G \in [0\;\;1]} U_i(\gamma, \G_{-i}) .$
 
\begin{figure*}[!h]
	\centering
\hspace{-.5cm}	\subfloat[]{\includegraphics[scale=0.24]{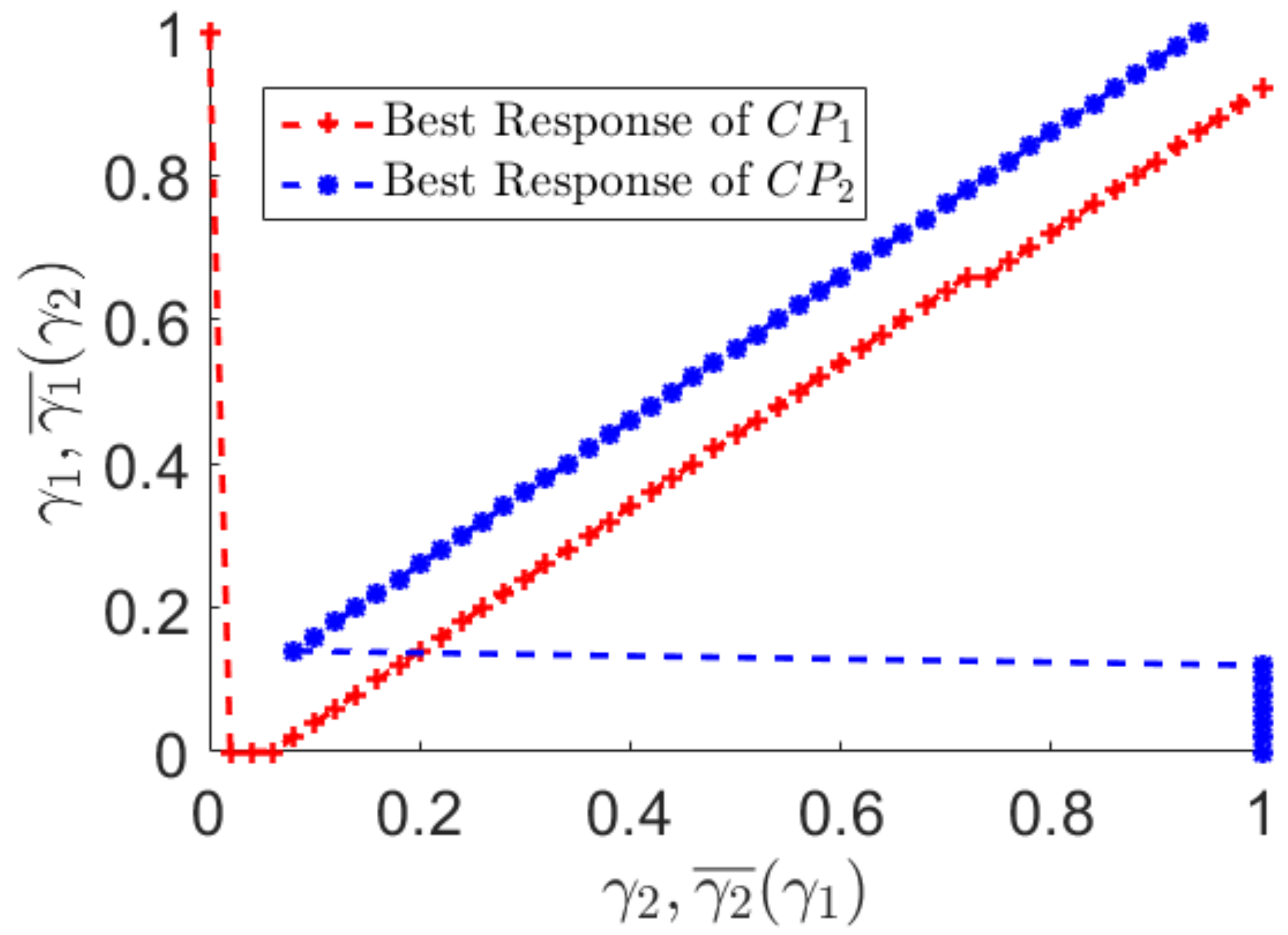}%
		\label{fig:BR1}}
		\subfloat[]{\includegraphics[scale=0.245]{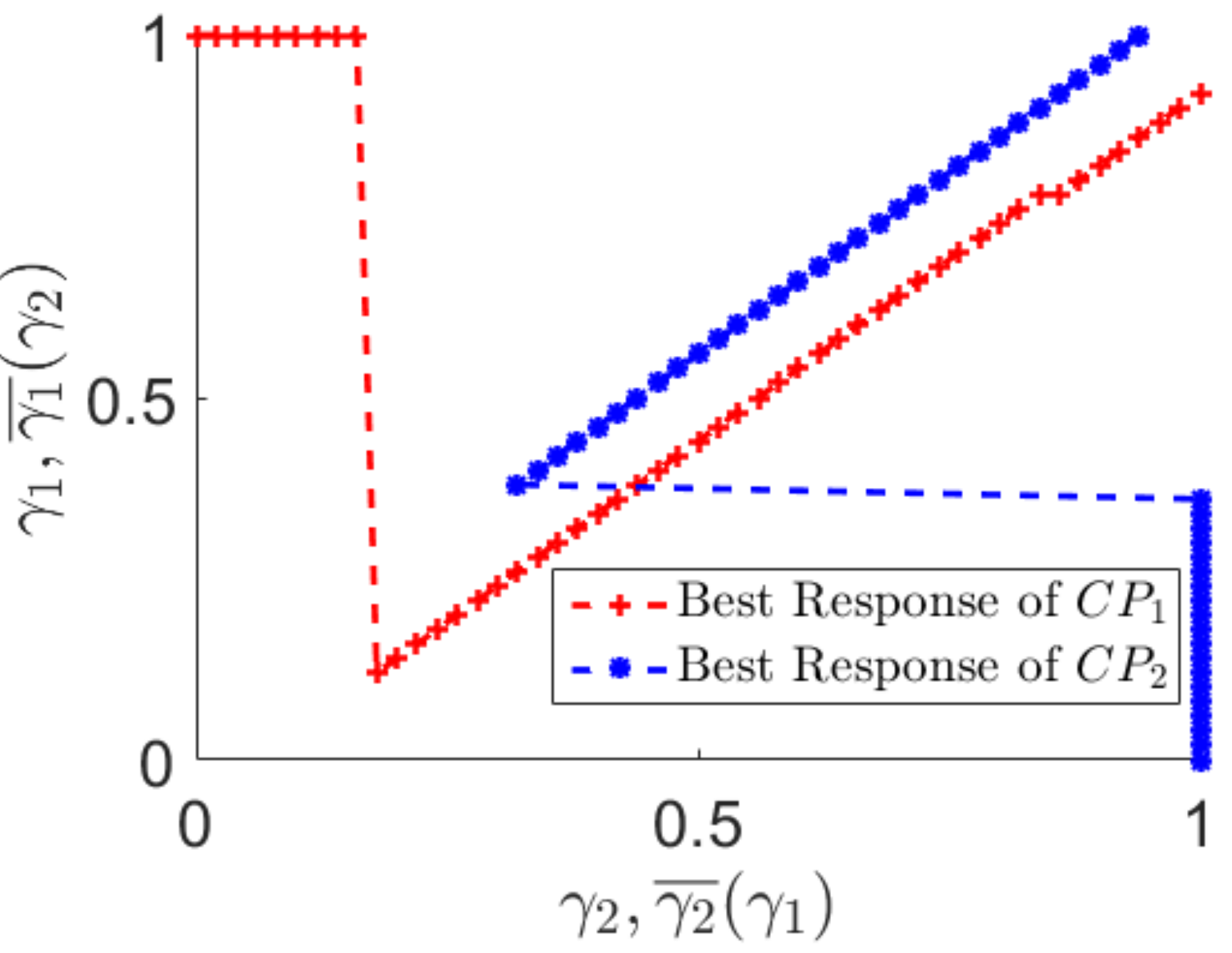}%
			\label{fig:BR2}}
		\subfloat[]{\includegraphics[scale=0.24]{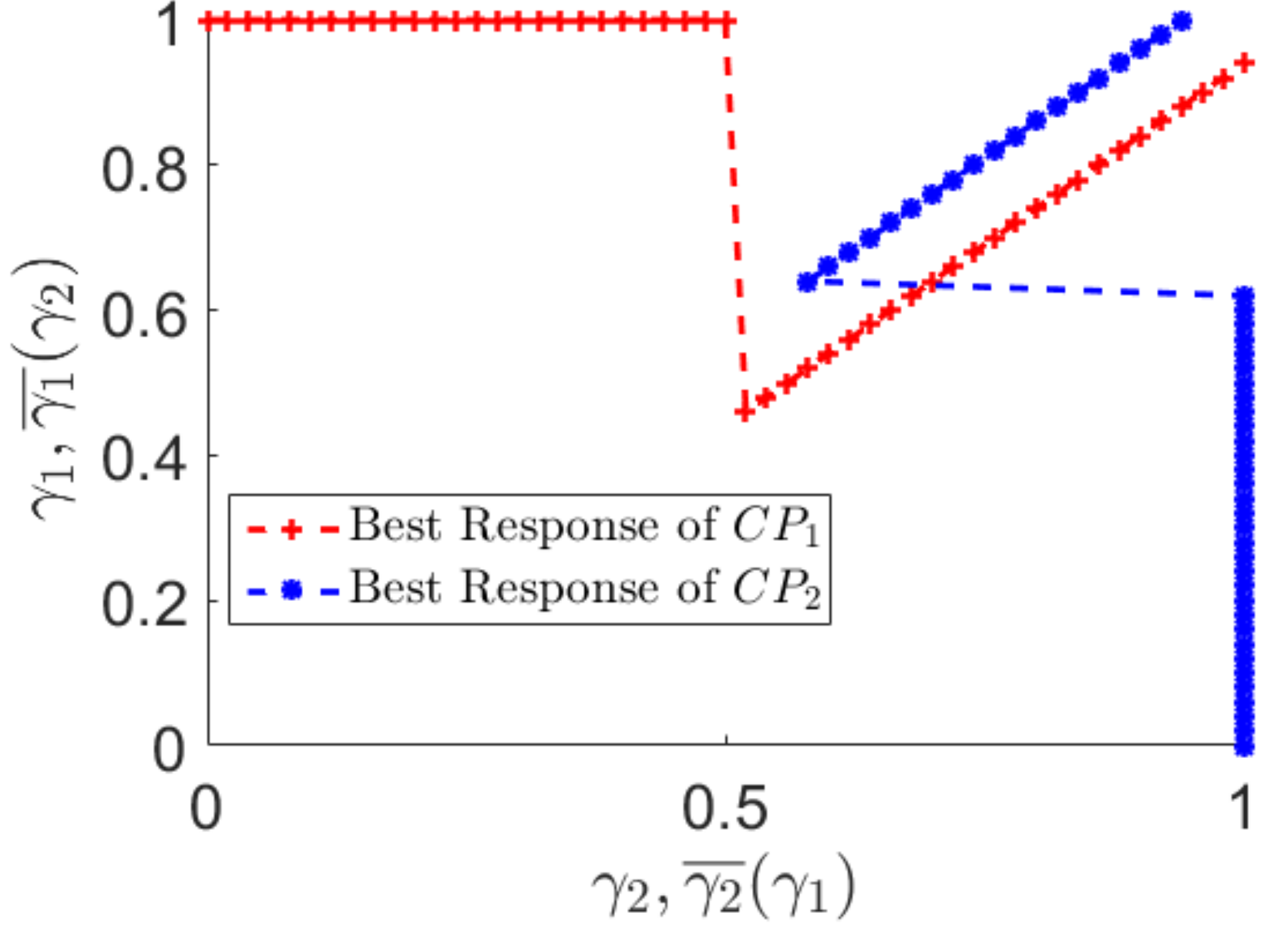}%
			\label{fig:BR3}}
		\subfloat[]{\includegraphics[scale=0.24]{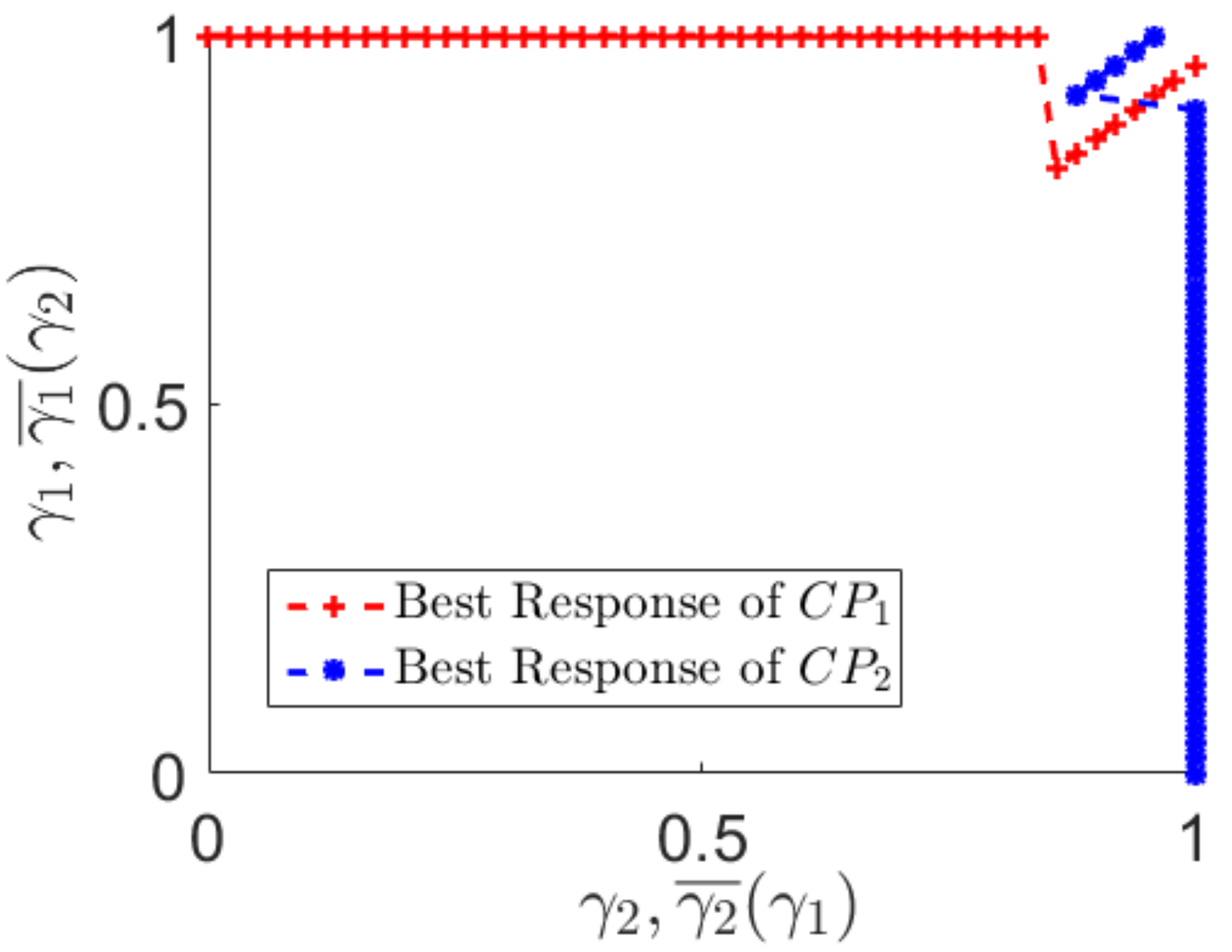}%
			\label{fig:BR4}}
			\caption {Best responses of both the CPs are shown with parameters $M_1=700, M_2=900, c=0.5, \rho=0.9, \beta=1$. We set $\lambda=1200, 1300, 1400, 1500$  in Figs. \ref{fig:BR1}, \ref{fig:BR2}, \ref{fig:BR3}, and \ref{fig:BR4}, respectively. In all both sub figures, the best response curve do not cross each other, hence equilibrium does not exist. Note that the sharp drops denote the discontinuity in best response behavior. The values in vertical and horizontal drops (dotted lines) does not correspond to range of the BR curves.}
			\label{fig:BR}
		\end{figure*}

The utility functions are non-linear as illustrated in Figs. \ref{fig:UtilityComp}. In Fig. \ref{fig:Utility1}, we depict utility functions of $CP_1$ and $CP_2$ as a function of $\G_1$ for a fixed value of $\G_2$. Note that $CP_1$ has multiple optimum points. Similar behavior is observed in Fig. \ref{fig:Utility2}. Further, notice that utilities of $CP_1$ and $CP_2$ have a steep slopes as $\G_1 \rightarrow \G_2$ and $\G_2 \rightarrow \G_1$, respectively. This property of CP utilities give raise to discontinuity in their best response behavior. The discontinuity in the best response behavior leads to non-existence of Nash equilibrium as illustrated In Figs. \ref{fig:BR1} - \ref{fig:BR4}. As seen, they have a point of discontinuity and best response functions do not intersect. Hence equilibrium do not exists in these examples. In the following we restrict the actions of CPs to $\gamma_i \in \{0,1\}$, i.e., either sponsor or not-sponsor and study the non-cooperative game between the CPs.
%Before this, we address the impact of exogenous arrivals on the revenue of the ISPs.
% \begin{figure}[!h]
%	\begin{minipage}[t]{0.5\linewidth}
%		\centering
%		\includegraphics[scale=.2]{figures/BR1}
%		\caption{Best response curves}
%		\label{fig:BR1}
%	\end{minipage}
%	\hspace{0.cm}
%	\begin{minipage}[t]{0.5\linewidth} 
%		\centering
%		\includegraphics[scale=.2]{figures/BR2}
%		\caption{Best response curves}
%		\label{fig:BR2}
%	\end{minipage}
%The figures demonstrates that the best response curves need not interesect and hence
%equilibrium may not exits.        
%\end{figure}

%
%\begin{figure*}[!h]
%	\centering
%	\subfloat[]{\includegraphics[scale=0.26]{figures/Utilities1}%
%		\label{fig:Utility1}}
%	\hspace{2cm}
%	\subfloat[]{\includegraphics[scale=0.26]{figures/Utilities2}%
%		\label{fig:Utility2}}
%	\caption {Utility of $CP_1$ and $CP_2$ for $M_1=700, M_2=900, c=0.5, \rho=0.9, \beta=1, \lambda=1000$. In \ref{fig:Utility1} we set $\gamma_2=0.5$ and utilities are shown as a function of $\G_1$.   In \ref{fig:Utility2}, we set $\gamma_1=0.5$ and utilities are shown as a function of $\G_2$. The utilities show multiple stationary points.}
%	\label{fig:Utilities}
%\end{figure*}
% The best response satisfy the following property:

%\begin{proposition}
%	\label{prop:BRProperty}
%For all $i=1,2$, $\overline{\G}_i(\G_{-i})$ is strictly increasing in $\G_{-i}$.
%\end{proposition}
%\begin{proof}
%For each $i=1,2$, consider  ${\G}^\prime_{-i}\geq\G_{-i}$. From Lemma (\ref{lma:EquRateMonotone}) we have $\lambda_i^*(\G_i, \G^\prime_{-i})\geq \lambda_i^*(\G_i, \G_{-i})$. Hence, $U_i(\G_i, \G^\prime_{-i})\geq U_i(\G_i, \G_{-i})$ for all $\G_i$. We thus conclude that $\overline{\G}_i(\G_{-i})$ is monotonically increasing in $\G_{-i}$.
%\end{proof}
%
%\begin{thm}
%Nash equilibrium for the non-cooperative game among the CPs exists and is unique.
%\end{thm}
%\begin{proof}
%The uniqueness property follows from the strict monotonicity property of the best response given in Proposition (\ref{prop:BRProperty}).
%\end{proof}
% \begin{figure}
% \centering
%  \vspace{-1cm}
% \includegraphics[scale=.2]{figures/CPUtilityComp}
% \vspace{-1cm}
% \end{figure}

%% file: ExogenousArrivals.tex
In the previous setup the total traffic rate is a constant and the total revenue for the ISP is $\lambda c$ irrespective of the subsidy set by the CPs. This makes the ISP indifferent to the differential pricing and zero-rating strategy proposed by the CPs. However, higher subsidy by the CPs may attract more traffic and in turn result in higher demand and then potential revenue for the ISP. To account for this, we next consider a model with exogenous generation of traffic in addition to the usual traffic of rate $\lambda$. The exogenous traffic corresponds to the increased  demand that the CPs attract, in addition to their usual traffic, by offering subsidies which increase the total traffic generated by end users.  
Naturally, higher is the subsidy offered by an CP higher is the exogenous traffic it attracts. 
We model the exogenous traffic generated for $CP_i$ offering subsidy $\gamma_i$ to be linear in $\gamma_i$  and given by $\lambda_0 (1-\gamma_i)$, where $\lambda_0 \geq 0 $ is a fixed constant which represents the global maximum demand. The total traffic $\tilde{\lambda}_i$ for $CP_i, i=1,2$, when it attracts the usual traffic $\lambda_i$ and offers subsidy $\gamma_i$, is then determined by: $\tilde{\lambda}_i=\lambda_i + \lambda_0 (1-\gamma_i)$, where $\lambda_1+\lambda_2 =\lambda$, and the total traffic in the networks is given by $\tilde{\lambda}:=\tilde{\lambda}(\gamma_1, \gamma_2)=\lambda + \sum_{i=1}^2 \lambda_0 (1-\gamma_i)$.
We continue to make the assumptions $A1-A3$ as earlier. Note that with exogenous arrivals it may happen that $\sum_{i=1}^2 m_i < \tilde{\lambda}$ if the subsidy offered by the CPs is too high hence restraining the CPs from taking such actions.

The usual traffic $\lambda$ gets split between the CPs competitively by the end users based on the QoS experienced and the access price, whereas the exogenous traffic is fixed and depends only on the subsidy factory. At equilibrium, the total traffic at $CP_i$ is therefore given by $\tilde{\lambda}_i^*=\lambda_i^* + \lambda_0(1-\gamma_i)$ where $(\lambda_1^*, \lambda_2^*)$ are equilibrium flows set according to Wardrop criteria as earlier.

%where $\tilde{\lambda}_i^*=\lambda_i^* + \lambda_0(1-\gamma_i)$
With exogenous arrivals, mean delay defined in Section \ref{sec:MeanDelay} can be redefined as
\begin{equation}
D(c,\gamma_1,\gamma_2)=\sum_{i=1}^{N}\frac{\tilde{\lambda}_i^*}{\tilde{\lambda}}\frac{1}{m_i- \tilde{\lambda}_i^*}.
\end{equation}
%and the utility of $i$th CP is given as 
%\begin{equation}
%	U_i(\gamma_i, \gamma_{-i})=(\beta- (1-\gamma_i)\rho c)\tilde{\lambda}_i^*
%\end{equation}
The analysis of mean delay and the results derived in the previous sections remain the  same after replacing the capacities $m_1$ and $m_2$ by $m_1 -\lambda_0 (1-\gamma_1)$ and $m_2-\lambda_0(1-\gamma_2)$, respectively, and $\sum_{i=1}^2 m_i \leq \tilde{\lambda}$.

\noindent
The utilities of CPs with exogenous arrivals are redefined as
\begin{equation}
\label{eqn:CPUtility_Exog}
U_i(\G_i, \G_{-i})=(\beta- (1-\G_i)\rho c)\tilde{\lambda}_i^* \quad \forall \; i=1,2.
\end{equation}
The condition for $CP_1$ to earn higher revenue than that of $CP_2$ under exogenous arrivals remains the same as in Proposition  \ref{prop:CP1larger} after replacing $\lambda$ by $\tilde{\lambda}$ and $\lambda_1^*$ by $\tilde{\lambda}_1^*$. Hence, our earlier observation that the CP with lower QoS may earn higher revenue that of the CP offering higher QoS holds under exogenous arrivals.
%
%\begin{proposition}
%Let $\overline{\beta}=\rho c/\beta\leq 1$. For a given $\gamma_2, \gamma_1\geq 0$, $U_1(\gamma_1,\gamma_2) \geq U_2(\gamma_2,\gamma_1)$ iff
%\begin{equation}
%\lambda_1^* \geq\frac{ \lambda+2\lambda_0-\lambda_0(\gamma_1+\gamma_2)}{\frac{1-(1-\gamma_1)\overline{\beta}}{1-(1-\gamma_2)\overline{\beta}}+1}-\lambda_0(1-\gamma_1).
%\end{equation}
%\end{proposition}
%Now, we analyze the utility function for the non cooperative game between the CPs for existence of Nash Equilibrium. 
Similar to previous model, the best response for CPs with exogenous traffic also have discontinuities and leads to non-existence of pure strategy Nash equilibrium.

\noindent
{\bf Elastic Demand:} One could also model the total demand generated by the users as elastic in the subsidy factor and model it as $\Lambda+\Lambda_0(1-\gamma_1)+\lambda_0(1-\gamma_1)$, where $\Lambda$ and $\Lambda_0$ are constants, which gets competitively split across the CPs. The difference between this model compared to exogenous arrivals is that here subsidy of each CP's effects the total demand not just its demand. The analysis with the elastic model remains essentially same as that in Sections \ref{sec:CPBehavior} and \ref{sec:CPGame}. Hence we skip the details. In the subsequent models we only consider the model with exogenous arrivals.

%% file: CPGame.tex
In this section we focus on the study of non cooperative game where CPs only play either Sponsor (S) or Not-sponsor (N) actions, i.e., $\G_i \in \{0,1\} $ for all $i=1,2$ (no partial sponsorship is allowed). The four possible action profiles are denoted as $(S,S), (N,N), (S,N),$ and $(N,S)$. Here, profile $(S,N)$ indicates that action of $CP_1$ is $S$ (sponsor) and that of $CP_2$ is $N$ (non-sponsor). For ease of notation, we add subscripts $00, 11, 01, 10$ to the minimum equilibrium utility ($\alpha$) of these action profiles, respectively. The following theorem gives equilibrium rates associated with these action profiles. In the following we assume that $\overline{m} > 2 \lambda_0$, i.e., the excess capacity is enough to support all of the exogenous traffic and $m_i > \lambda_0\; \forall i$, i.e., each CP has enough capacity to handle all of the exogenous traffic it can attract.

\begin{thm}
	\label{thm:EquRatesSpecialCases}
	The equilibrium rates are described as follows:
	\begin{enumerate}
		\item For action profile $(S,S)$:
                     \\ $\tilde{\lambda_i^*}=m_i-\lambda_0-\frac{1}{\alpha_{00}} \;\; \forall i$  where $ \alpha_{00}=\frac{2}{\overline{m}-2\lambda_0}.$
		\item For action profile $(N,N)$:
		\\$\tilde{\lambda_i^*}=m_i-\frac{1}{\alpha_{11}-c} \;\; \forall i$ where $\alpha_{11}=c+\frac{2}{\overline{m}}$.
		\item For action profile $(S,N)$:
		\\$\tilde{\lambda_1^*}=m_1-\lambda_0-\frac{1}{\alpha_{01}}$ and
		$\tilde{\lambda_2^*}=m_2-\frac{1}{\alpha_{01}-c}$ where $\alpha_{01}=c/2+1/(\overline{m}-\lambda_0)+\sqrt{c^2/4+ 1/(\overline{m}- \lambda_0)^2}$.
		\item For action profile $(N,S)$:
		\\$\tilde{\lambda_i^*}=m_1-\frac{1}{\alpha_{10}-c}$ and $\tilde{\lambda_2^*}=m_2-\lambda_0-\frac{1}{\alpha_{10}} $ where $\alpha_{10}=\alpha_{01}$. %\hfill\IEEEQED
	\end{enumerate}
\end{thm}

\begin{cor}
	\label{corol:MinCostRelations}
	The minimum equilibrium cost under different action profiles satisfy the following properties:
	\begin{itemize}
		\item $\alpha_{00}\geq \alpha_{11}-c$,
		\item $\alpha_{01}\geq \alpha_{00}\geq \alpha_{10}-c$,
		\item $c+ 1/(\overline{m}-\lambda_0) \leq \alpha_{01}\leq c+ 2/(\overline{m}-\lambda_0)$. 	\end{itemize}
\end{cor}
Considering that each CP play only pure actions $\{\textit{S,N}\}$, the non-cooperative setting is a described by a $2\times 2$ matrix game between the CPs. The utilities for both CPs over actions $\{\textit{S,N}\}$ are given in Table 1. 
%\begin{table}
%	\begin{center}
%		\begin{tabular}{|c|c|c|}\hline
%			$ A $ & $S $ & $ N $ \\ \hline
%			S & $U_1=(\beta-\rho c)(m_1-\lambda_0-1/\alpha_{00})$ & $U_1=(\beta-\rho c)(m_1-\lambda_0-1/\alpha_{01})%$ \\
%			$ $ & $U_2=(\beta-\rho c)(m_2-\lambda_0-1/\alpha_{00})$ & $U_2=\beta(m_2-\lambda_0-1/(\alpha_{01}-c))$ \%\ \hline
%			N & $U_1=\beta(m_1-\lambda_0-1/(\alpha_{10}-c))$ & $U_1=\beta(m_1-\lambda_0-1/(\alpha_{11}-c))$ \\
%			$ $ & $U_2=(\beta-\rho c)(m_2-\lambda_0-1/\alpha_{10})$ & $U_2=\beta(m_2-\lambda_0-1/(\alpha_{11}-c))$ \%\ \hline
%		\end{tabular}
%	\end{center}
%	\caption{Utility for CPs over actions $\{S,N\}$}  \label{tab:Utilities}
%\end{table}

\begin{table}
	\begin{center}
		\begin{tabular}{|c|c|c|}\hline
			$ \mbox{Action} $ & $S $ \\ \hline
			S & $U_1=(\beta-\rho c)(m_1+\lambda_0-1/\alpha_{00})$ \\
			$ $ & $U_2=(\beta-\rho c)(m_2+\lambda_0-1/\alpha_{00})$ \\ \hline
			N & $U_1=\beta(m_1+\lambda_0-1/(\alpha_{10}-c))$ \\
			$ $ & $U_2=(\beta-\rho c)(m_2+\lambda_0-1/\alpha_{10})$\\ \hline
		\end{tabular}
	\end{center}
    \begin{center}
		\begin{tabular}{|c|c|c|}\hline
			$  \mbox{Action}  $ &$ N $ \\ \hline
			S & $U_1=(\beta-\rho c)(m_1+\lambda_0-1/\alpha_{01})$ \\
			$ $ & $U_2=\beta(m_2+\lambda_0-1/(\alpha_{01}-c))$ \\ \hline
			N & $U_1=\beta(m_1+\lambda_0-1/(\alpha_{11}-c))$ \\
			$ $ & $U_2=\beta(m_2+\lambda_0-1/(\alpha_{11}-c))$ \\ \hline
		\end{tabular}
	\end{center}
	\caption{Utility for CPs over actions $\{S,N\}$}  \label{tab:Utilities}
\end{table}

\begin{figure}[!h]

	\centering
	\includegraphics[scale=.245]{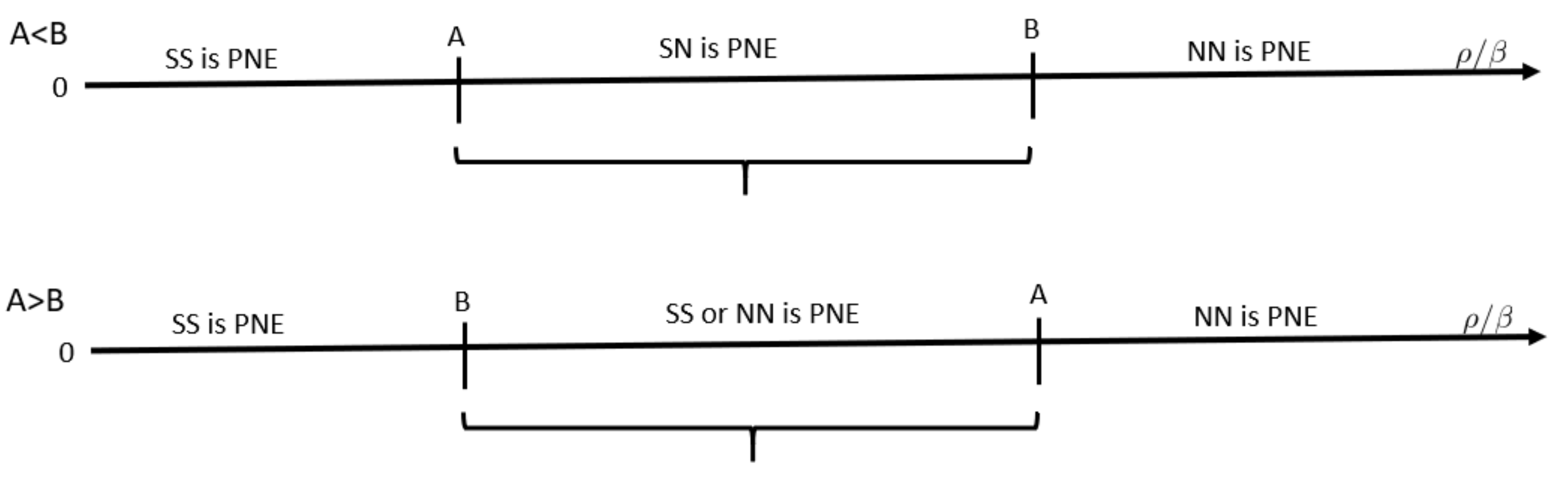}
	
	\caption{The PNE as a function of $\rho/\beta$.}
		\label{fig:PNE1}
\end{figure}

Though it is well known that a mixed Nash equilibria always exists in a finite matrix game but pure Nash equilibria may not exists\footnote{Though $\gamma_i \in  [0,1]$ can be thought of as probability distribution over $\gamma_i\in \{0,1\}$, it is to be noted that our earlier observation that non existence of Nash equilibrium does not contradict the fact mixed Nash equilibrium always exits. The CP utilities are highly nonlinear in $\gamma_i$s and expected utility over any mixed equilibria dervied from utilities associated with pure actions does not have the same form as in (\ref{eqn:CPUtility_Exog}). }. Since the decision of CPs has to be deterministic, we are interested in the study of Pure strategy Nash Equilibria (PNE) and its properties. The following theorem characterizes all possible PNE.  
\begin{thm}
	\label{thm:PNE}
Let $\rho, c, \beta$ be given. Then,
\begin{itemize}
	\item $(S,S)$ is a PNE if and only if  
	\begin{equation}
	\label{eqn:SS}
	\rho/\beta\leq \frac{ 1/(\alpha_{10}-c)+\lambda_0-1/\alpha_{00}}{c(m_2+\lambda_0-1/\alpha_{00})}:=A
	\end{equation}
	\item $(N,N)$ is a PNE if and only if 
	\begin{equation}
		\label{eqn:NN}
\rho/\beta \geq \frac{1/(\alpha_{11}-c)+\lambda_0-1/\alpha_{01}}{c(m_1+\lambda_0-1/\alpha_{01})}:=B
	\end{equation}
	\item $(S,N)$ is a PNE if and only if
	\begin{eqnarray}
		\label{eqn:SN}
	\frac{ 1/(\alpha_{10}-c)+\lambda_0-1/\alpha_{00}}{c(m_2+\lambda_0-1/\alpha_{00})} \leq \rho/\beta \nonumber \\
	 \rho/\beta \leq\frac{1/(\alpha_{11}-c)+\lambda_0-1/\alpha_{01}}{c(m_1+\lambda_0-1/\alpha_{01})}
	\end{eqnarray}
	\item $(N,S)$ cannot be a PNE.
\end{itemize}
\end{thm}
%
%\begin{cor}
%	\label{corol:NoPNE}
%	For all $c>0$, we have $C>D$ and hence $(N,S)$ cannot be a PNE.
%\end{cor}
Theorem \ref{thm:PNE} characterizes all PNE and Figure \ref{fig:PNE1} depicts all possible PNE in different range over $\rho/\beta$ .

%\begin{figure*}[!h]
%	\centering
%		\subfloat[]{\includegraphics[scale=0.05]{figures/AB1}%
%		\label{fig:AB1}}
%	\hspace{0.6cm}
%	\subfloat[]{\includegraphics[scale=0.05]{figures/AB2}%
%		\label{fig:AB2}}
%		\caption {The comparision $A$ and $B$ with respect to change in parameters $c$ in Figs. \ref{fig:AB1} and \ref{fig:AB2} for different values of $m_1$.   We set $\beta=1$,$\rho=.9$, $\lambda=1200$, $\lambda_0=200$ and $m_2=1000$ throughout. In Fig. \ref{fig:AB1}, we set $m_1=900$ and in Fig. \ref{fig:AB2}, we set $m_1=650$.}
%	\label{fig:Gain}
%\end{figure*}

%
%We also continue to make Assumptions A1-A3. For convenience, we re-define $\overline{m}$ as $\overline{m}:=\sum_{i=1}^2 m_i -\tilde{\Lambda}=\sum_{i=1}^2 \tilde{m}_i -\Lambda$ where  $\tilde{m}_i=m_i-\Lambda_0(1-\gamma_i)$.The elastic traffic $\tilde\Lambda$ get splitted between CPs compettitively by the users based on the QoS experienced and the access price say $(\tilde\Lambda_1^*, \tilde\Lambda_2^*)$  which are set according to the Wardrop equilibrium as earlier as is given by
%\begin{equation}
%\tilde\Lambda_i^*=\Lambda_i+\Lambda_0(1-\gamma_i)=m_i-\frac{1}{\alpha-\gamma_i c} \forall i=1, 2
%\end{equation}
%
%where $\alpha:=\alpha(\gamma_1,\gamma_2)$ is the equilibrium cost given by
%\begin{equation}
%\alpha=\frac{c(\gamma_1+\gamma_2)}{2}+\frac{1}{\overline{m}}+ \sqrt{\frac{(c(\gamma_1-\gamma_2))^2}{4}+\frac{1}{\overline{m}^2}}.
%\end{equation}
%
%
%Similar to exogenous arrival model, we can define mean delay for end users and Utility of CPs. With these new definitions, all the results and properties obtained are similar to the exogenous arrival model. Infact, Non-cooperative game between CPs shows exactly same behaviour as that of exogenous model.

%% file: RGF.tex
In this section we study the revenue gain for ISP induced by applying differential pricing mechanism. The exogenous traffic generated by the end users increases with higher subsidy proposed by the CPs, which induces higher revenue for the ISP. To measure this revenue gain, we consider the metric called 'Revenue Gain Factor' (RGF) defined as the ratio of ISP revenue under the differential pricing scheme and that under the neutral regime where none of the CPs subsidize the access price, i.e., $\gamma_i=1$ for any CP $i$, then:

\begin{figure*}[!tb]
	\centering
	\subfloat[]{\includegraphics[scale=0.06]{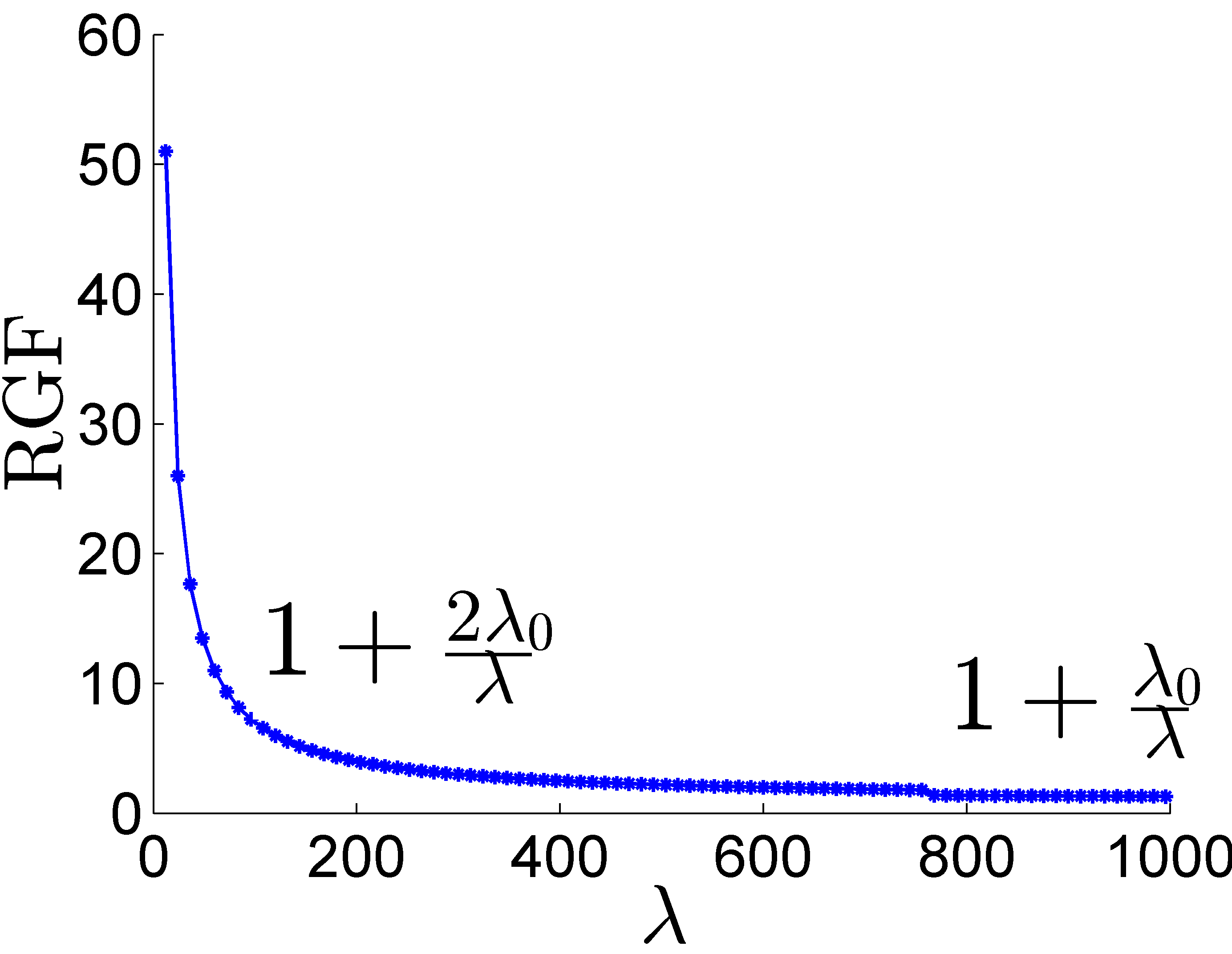}%
		\label{fig:Gain1}}
	\hspace{0.4cm}
	\subfloat[]{\includegraphics[scale=0.37]{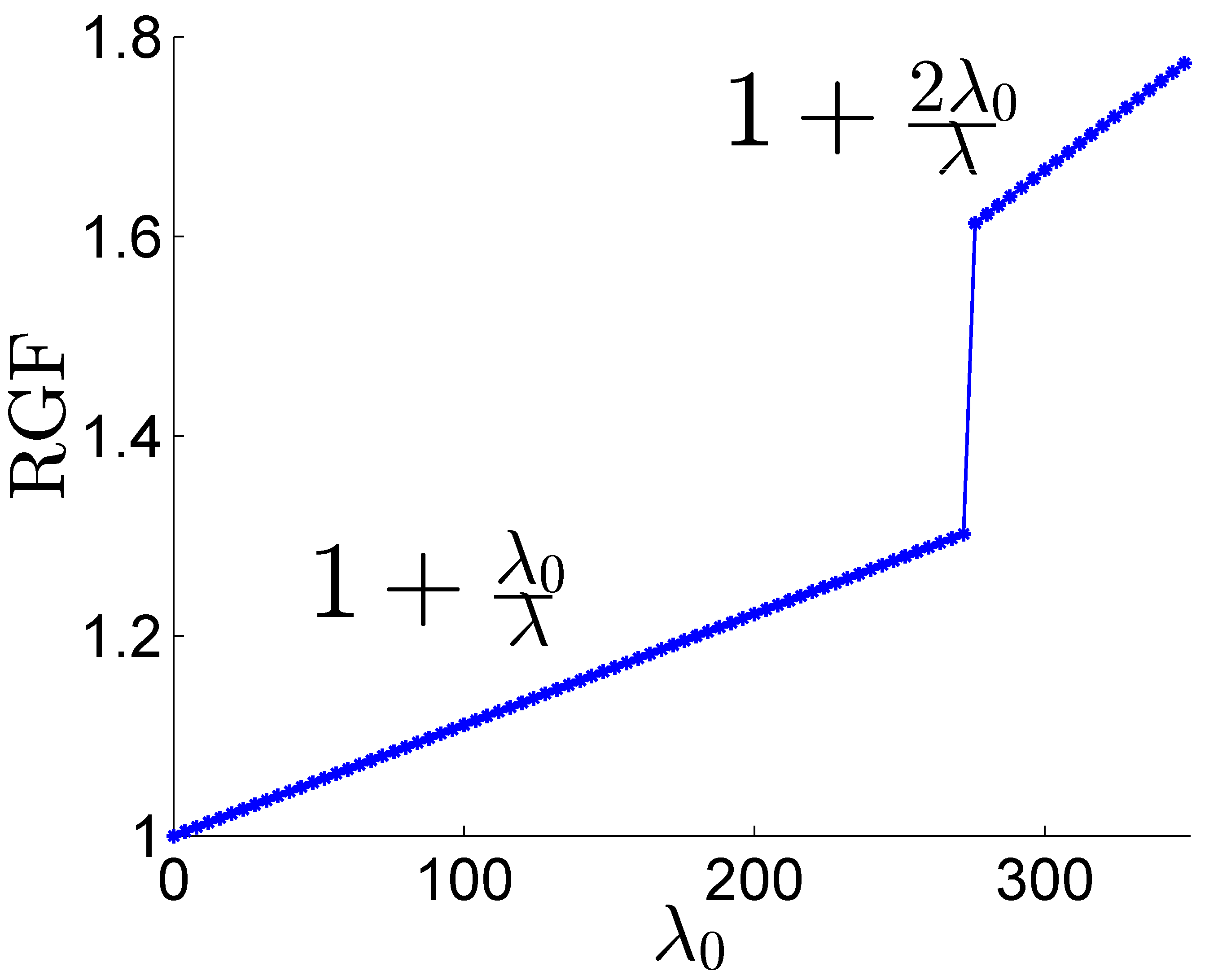}%
		\label{fig:Gain2}}
	\hspace{0.4cm}
	\subfloat[]{\includegraphics[scale=0.38]{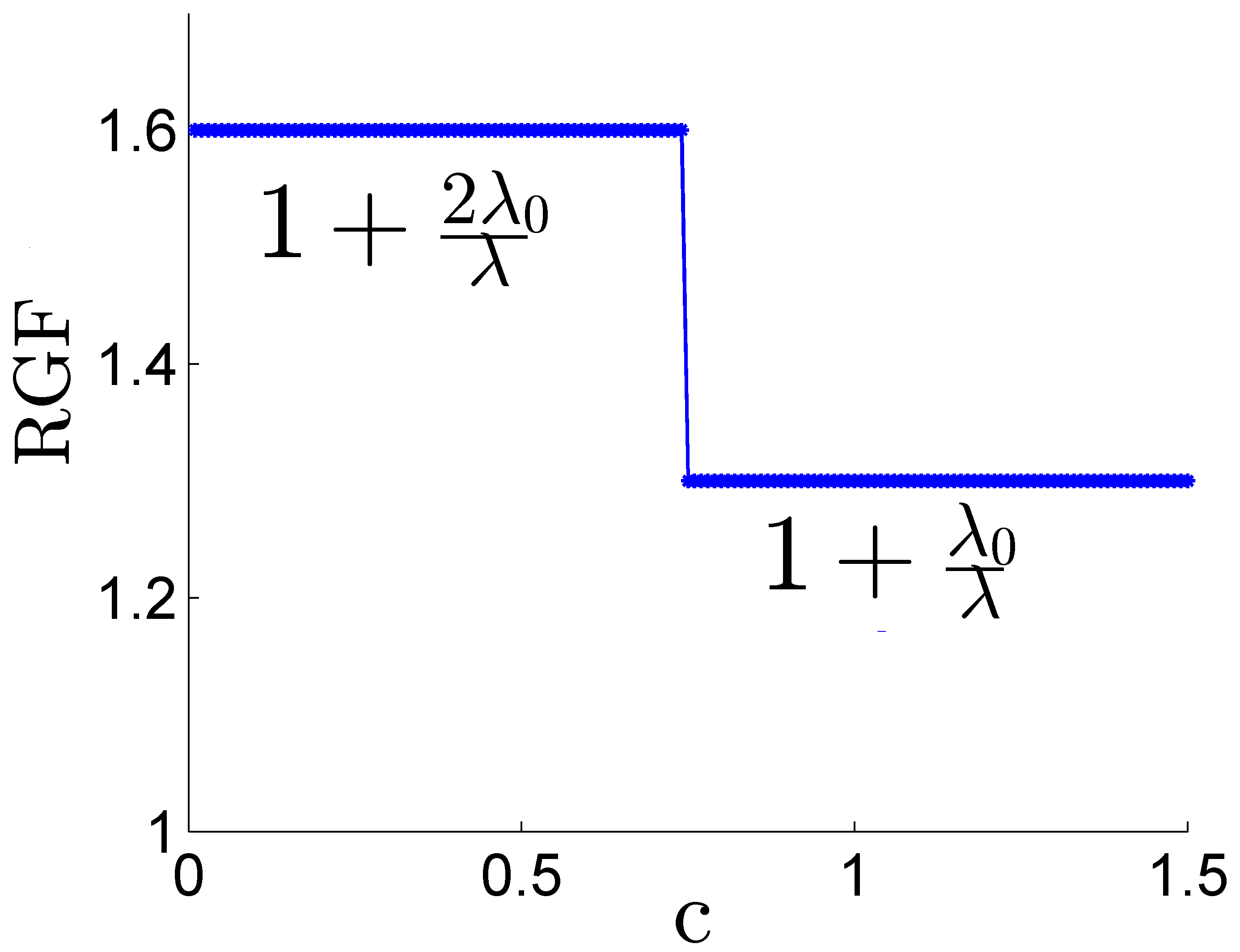}%
		\label{fig:Gain3}}
	\caption {The change in 'Revenue Gain Factor' for ISP and shift in PNE point is being studied with respect to change in parameters $\lambda$, $\lambda_0$ and $c$ in Figs. \ref{fig:Gain1}, \ref{fig:Gain2} and \ref{fig:Gain3} respectively.   We set $\beta=1$, $\rho=.9$, $m_1=700$ and $m_2=900$ throughout. In Fig. \ref{fig:Gain1}, we set $c=0.7$ and $\lambda_0=300$, in Fig. \ref{fig:Gain2}, we set $c=0.5$ and $\lambda=1000$ and in Fig. \ref{fig:Gain3}, we set $\lambda=100$ and $\lambda_0=300$.}
	\label{fig:Gain}
\end{figure*}

\begin{multline}
RGF=\frac{(\lambda+\lambda_0(1-\gamma_1)+\lambda_0(1-\gamma_2))\rho c}{\lambda \rho c} \\=1+\frac{\lambda_0(1-\gamma_1)+\lambda_0(1-\gamma_2)}{\lambda}.
\end{multline}

The following proposition describes explicit value of RGF depending on the action profile of CPs.
\begin{proposition}
	\label{thm:RGF}
	RGF under different action profiles are as follows:
	\begin{itemize}
		\item For action profile $(S,S)$:
		$RGF=1+\frac{2\lambda_0}{\lambda}$
		\item For action profile $(N,N)$:
		$RGF=1$, i.e., there is no revenue gain for the ISP
		\item For action profile $(S,N)$:
		$RGF =1+\frac{\lambda_0}{\lambda}$
		\item For action profile $(N,S)$:
		$RGF=1+\frac{\lambda_0}{\lambda}.$%\hfill\IEEEQED
	\end{itemize}
\end{proposition}
Obviously, the RGF is the highest when both CPs sponsor. However, the action profile $(S,S)$ may not be always a pure Nash equilibrium as proved in previous section. In the following we illustrate the behavior of the RGF metric at equilibrium with respect to parameters $\lambda$, $\lambda_0$ and $c$. We note that the value of RGF at equilibrium depends on $c$ through equilibrium values of $\gamma_i^*, i=1,2$.  From Fig. \ref{fig:Gain1} it is observed that as $\lambda$ increases, RGF initially decreases steeply and then at a slower rate. This change is due to shift of PNE point from $(S,S)$ to $(S,N)$ when increasing $\lambda$. This observation is natural -- if the usual traffic is already high, then the additional traffic from subsidy may not improve the ISP revenue much. Thus the ISP prefers a differential pricing regime only in the case where the intrinsic traffic is not significant compared to the exogenous traffic.  Fig. \ref{fig:Gain2} describes RGF as a function of $\lambda_0$ for a fixed $\lambda$. Initially the gain increases at smaller rate with increase in exogenous traffic ($\lambda_0$) and later the rate increases. This shift in behavior is again due to shifting PNE points from $(S,N)$ to $(S,S)$ as CPs are likely to go for full sponsorship if more exogenous traffic is generated. Thus, higher the exogenous traffic, more is the the ISP's preference for the differential pricing.  Fig. \ref{fig:Gain3} shows that RGF is unaffected by the change in cost keeping PNE point fixed. However, with the increase in cost, PNE shifts from $(S,S)$ to $(S,N)$ and thereby causing the drop in RGF. This implies fixing higher access price may not be beneficial for ISP under exogenous arrival of demand.

%% file: MultiISP.tex
\begin{figure}
	\centering
	\includegraphics[scale=.5]{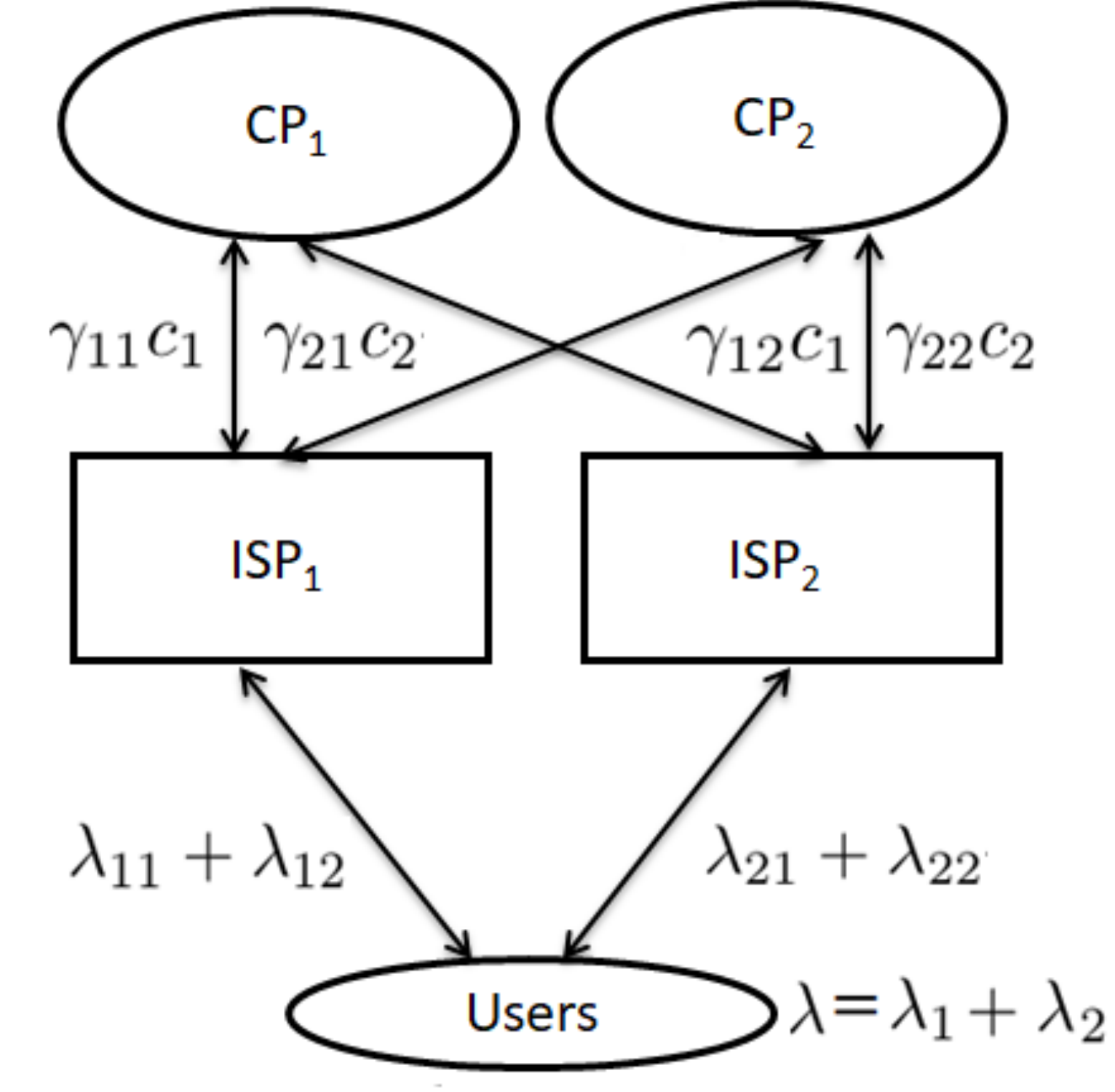}
	\caption{ Interaction between the Multi-ISP, CPs and end users.}
		\label{fig:Connections1}
\end{figure}

In this section, we extend our analysis for the case of oligopoly market with several ISPs. We consider two ISPs but the analysis can be extended to more than two. Each ISP connects both the CPs to the end users and set the access price independently. Knowing the access price, the CPs set the subsidy factor for each ISP independently. Let $c_i, i=1,2$ denotes the access price set by $ISP_i$  and $\gamma_{ij}$ denote the subsidy factor set by $CP_j$ for traffic generated through $ISP_i$. We denote the traffic rate that flows to $CP_j$ from $ISP_i$ as $\lambda_{ij}$. The exogenous traffic rate from $ISP_i$ to $CP_j$ is $\lambda_0(1-\gamma_{ij})$. Without loss of generality we assume that $c_1<c_2$. The interaction framework is described in figure \ref{fig:Connections1}.

The cost of an end user that connects via  $ISP_i$ to $CP_j$ while it is receiving traffic at rate $x$ is given by:
\begin{equation}
\label{eqn:MultiISPUserUtility}
C_{ij}(x):=C_{ij}(x,m_j,c_i,\gamma_{ij})=\begin{cases}
\frac{1}{m_j -x}  + \gamma_{ij}c_i &\mbox{  if } m_j > x, \\
\infty &\mbox{otherwise.}
\end{cases}
\end{equation}

\noindent
Let $\lambda_{ij}^*$ denote the flows at equilibrium for all $i,j=1,2$. 
\begin{lemma}
	\label{lma:RatesMultiISP}
For a given $\gamma_{ji}$ and $c_i, \forall \;i,j=1,2$, we have
\begin{itemize}
	\item $\lambda^*_{1j}>0$ and $\lambda_{2j}^*=0$ if and only if $\gamma_{1j}c_1 < \gamma_{2j}c_{2}$,
	\item $\lambda_{1j}^*=0$ and $\lambda_{2j}^*>0$ if and only if $\gamma_{1j}c_1 > \gamma_{2j}c_2$,
	\item $\lambda_{1j}^*=\lambda_{2j}^*$ if and only if $\gamma_{1j}c_1 = \gamma_{2j}c_2$.
\end{itemize}

\end{lemma} 

As in Section \ref{sec:CPGame}, we assume that CPs either Sponsor or Not Sponsor traffic of the ISPs, i.e. $\gamma_{ij} \in \{0,1\}, \forall i,j$. This decision is based on the access price proposed by ISPs. Further, CPs are assumed to sponsor only traffic on one ISP only, not both. Indeed, a CP cannot contract with two different ISPs. The resulting action profile for CPs is then $\{SN, NS, NN\}$, where action $SN$ denote that traffic form $ISP_1$ is sponsored while that from $ISP_2$ is not sponsored, $NS$ denotes that the other way around and $NN$ denotes that traffic from none of the ISPs is sponsored. 

 The $CP_j$'s utility is given by:
$$
U_j(\gamma_{ij})=\sum_i(\beta_j-\rho(1-\gamma_{ij})c_i)( \lambda_{ij}^*+\lambda_0(1-\gamma_{ij})).
$$ 
%where with abuse of notations we denote $\gamma_j:=(\gamma_{1j},\gamma_{2j})$. 
One can compute the equilibrium traffic $\lambda_{ij}^*$ each CP $j$ gets from each ISP $i$ using Wardrop conditions as earlier and the resulting utilities are summarized in Table \ref{tab:Utilities3}.

\begin{table}
	\begin{center}
		\begin{tabular}{|c|c|c|}\hline
			$\text{Action} $ & $SN $ \\ \hline
			$SN$ & $U_1=(\beta_{1}-\rho c_1)(m_1-\overline{m}/2+\lambda_0)$ \\
			$ $ & $U_2=(\beta_{2}-\rho c_1)(m_2-\overline{m}/2+\lambda_0)$ \\ \hline
			$NS$ & $U_1=(\beta_{1}-\rho c_2)(m_1-\overline{m}/2+\lambda_0)$ \\
			$ $ & $U_2=(\beta_{2}-\rho c_1)(m_2-\overline{m}/2+\lambda_0)$\\ \hline
			$NN$ & $U_1=\beta_{1}(m_1-1/(\alpha-c_1))$ \\
			$ $ & $U_2=(\beta_{2}-\rho c_1)(m_2-1/\alpha+\lambda_0)$\\ \hline
		\end{tabular}
	\end{center}
	
	\begin{center}
		\begin{tabular}{|c|c|c|}\hline
			$ \text{Action} $ & $NS$ \\ \hline
			$SN$ & $U_1=(\beta_{1}-\rho c_1)(m_1-\overline{m}/2+\lambda_0)$ \\
			$ $ & $U_2=(\beta_{2}-\rho c_2)(m_2-\overline{m}/2+\lambda_0)$ \\ \hline
			$NS$ & $U_1=(\beta_{1}-\rho c_2)(m_1-\overline{m}/2+\lambda_0)$ \\
			$ $ & $U_2=(\beta_{2}-\rho c_2)(m_2-\overline{m}/2+\lambda_0)$\\ \hline
			$NN$ & $U_1=\beta_{1}(m_1-1/(\alpha-c_1))$ \\
			$ $ & $U_2=(\beta_{2}-\rho c_2)(m_2-1/\alpha+\lambda_0)$\\ \hline
		\end{tabular}
	\end{center}
	
	\begin{center}
		\begin{tabular}{|c|c|c|}\hline
			$ \text{Action} $ & $NN$ \\ \hline
			$SN$ & $U_1=(\beta_{1}-\rho c_1)(m_1-1/\alpha+\lambda_0)$ \\
			$ $ & $U_2=\beta_{2}(m_2-1/(\alpha-c_1))$ \\ \hline
			$NS$ &  $U_1=(\beta_{1}-\rho c_2)(m_1-1/\alpha+\lambda_0)$ \\
			$ $ & $U_2=\beta_{2}(m_2-1/(\alpha-c_1)$ \\ \hline
			$NN$ & $U_1=\beta_{1}(m_1-\overline{m}/2)$ \\
			$ $ & $U_2=\beta_{2}(m_2-\overline{m}/2)$\\ \hline
		\end{tabular}
	\end{center}
	\caption{Utility for CPs over actions $\{SN,NS,NN\}$. Row actions correspond to $CP_1$ and column action to $CP_2$}  \label{tab:Utilities3}
\end{table}

%We could consider the following assumptions:
%\begin{itemize}
%\item each CP can make an agreement with only one ISP,
%\item only fully subsidizations are allowed, i.e. $\gamma_i^j \in \{N,S\}$.
%\end{itemize}
Utility of ISP $i$ is the total revenue earned from traffic that goes through his network, i.e.,
$$
R_i(c_i)=\sum_{j}\lambda_{ij}^*c_i.
$$
The following theorem describes the pure Nash equilibrium of the non-cooperative game between CPs, depending on main parameters of the model which are the access prices $c_i$ proposed by ISP $i$.
\begin{thm}
	\label{thm:PNE2}
Let $\rho$, $c$ and $\beta_j=\beta$ $\forall j$ given. Assume $c_1 < c_2$, then only $(SN,SN)$, $(NN,NN)$  and $(SN,NN)$ can be PNE. Specifically,
\begin{itemize}
\item  $(SN, SN)$  is the PNE if and only if  
\begin{equation*}
\label{eqn:SNSN}
\rho/\beta \leq \frac{1/(\alpha-c_1)-\overline{m}/2+\lambda_0}{c_1(m_2-\overline{m}/2+\lambda_0)},
\end{equation*}

\item  $(NN, NN)$ is the PNE if and only if 
\begin{equation*}
\label{eqn:NNNN}
\rho/\beta \geq \frac{\overline{m}/2-1/\alpha+\lambda_0}{c_1(m_1-1/\alpha+\lambda_0)},
\end{equation*}

\item $(SN, NN)$ is the PNE if and only if 
\begin{equation*}
\label{eqn:31}
\frac{1/(\alpha-c_1)-\overline{m}/2+\lambda_0}{c_1(m_2-\overline{m}/2+\lambda_0)}\leq \rho/\beta \leq \frac{\overline{m}/2-1/\alpha+\lambda_0}{c_1(m_1-1/\alpha+\lambda_0)}. 
\end{equation*}
\end{itemize}
The previous theorem suggests that CPs will not subsidize the traffic from ISP with higher access price. Further, coupled with Lemma \ref{lma:RatesMultiISP}, we note that traffic flow through ISP with higher cost is null at equilibrium.

%
% \begin{itemize}
%	\item  $(SN, SN)$  is a PNE if and only if  
%	\begin{equation}
%	\label{eqn:SNSN}
%	\rho/\beta \leq \frac{1/(\alpha-c_1)-\overline{m}/2}{c_1(m_2-\overline{m}/2)}
%	\end{equation}
%	\item  $(NN, NN)$ is a PNE if and only if 
%	\begin{equation}
%		\label{eqn:NNNN}
%\rho/\beta \geq \frac{\overline{m}/2-1/\alpha}{c_1(m_1-1/\alpha)}
%	\end{equation}
%           \item $(SN, NN)$ is a PNE if and only if 
%	\begin{equation}
%		\label{eqn:SNNN}
%\frac{1/(\alpha-c_1)-\overline{m}/2}{c_1(m_2-\overline{m}/2)}\leq \rho/\beta \leq \frac{\overline{m}/2-1/\alpha}{c_1(m_1-1/\alpha)} 
%	\end{equation}
% %         \item $(NN, SN)$ is a PNE if and only if 
%%	\begin{equation}
%%		\label{eqn:31}
%%\frac{1/(\alpha-c_1)-\overline{m}/2}{c_1(m_1-\overline{m}/2)}\leq \rho/\beta \leq \frac{\overline{m}/2-1/\alpha}{c_1(m_2-1/\alpha)} 
%%	\end{equation}
%\end{itemize}
\end{thm}
%
%We now consider exogenous demand model and assume that the exogenous traffic generated for $CP_i$ offering subsidy $\gamma^j_i$ to $ISP_j$ is linear in $\gamma^ j_i$ and is given as $\lambda_0 (1-\gamma^ j_i)$, where $\lambda_0 \geq 0 $ is a fixed constant. The total traffic for $CP_i, i=1,2$ through $ISP_j, j=1, 2$, when it attract the usual traffic $\lambda^j_i$ and offers subsidy $\gamma^j_i$  is given by $\tilde{\lambda}^j_i =\lambda^j_i + \lambda_0 (1-\gamma^j_i)$. The total traffic generated for $CP_1$ is given by $\tilde{\lambda}_1:=\tilde{\lambda}_1(\gamma^1_1, \gamma^2_1)=\lambda_1 + \sum_{j=1}^2 \lambda_0 (1-\gamma^j_1)$, and for $CP_2$ is given by $\tilde{\lambda}_2:=\tilde{\lambda}_2(\gamma^1_2, \gamma^2_2)=\lambda_1 + \sum_{j=1}^2 \lambda_0 (1-\gamma^j_2)$. Thus, total traffic generated in the network is $\tilde{\lambda}=\tilde{\lambda}_1+\tilde{\lambda}_2$. For this case, 
Let $RGF_i, i=1,2$ denotes the revenue gain factor for $ISP_i$ defined as follows:
\begin{eqnarray*}
RGF_i&=&\frac{(\sum_{j} \lambda_{ij}+\lambda_0(1-\gamma_{ij}))\rho c_i}{(\sum_j\lambda_{ij}) \rho c_i}, \\
&=&1+\frac{\sum_j\lambda_0(1-\gamma_{ij})}{\sum_j\lambda_{ij}}.
\end{eqnarray*}
%\begin{multline}
%RGF_2=\frac{(\lambda^2_1+\lambda^2_2+\lambda_0(1-\gamma^2_1)+\lambda_0(1-\gamma^2_2))\rho c_1}{(\lambda^2_1+\lambda^2_2) \rho c_1} \\=1+\frac{\lambda_0(1-\gamma^2_1)+\lambda_0(1-\gamma^2_2)}{\lambda^2_1+\lambda^2_2}
%\end{multline}
% The $CP_i$'s utility is:
%$$
%U_i(\gamma_i)=\sum_j(\beta_i^j-\rho(1-\gamma_i^j)c_j).
%$$ 
Considering exogenous arrivals, the usual traffic $\lambda$ gets split between the CPs competitively based on the QoS experienced and the access price whereas the exogenous traffic is fixed and depends only on the subsidy factory. The total traffic at $CP_j$ at equilibrium is given by $\tilde{\lambda}^*_j=\sum_i \lambda_{ij}^*+\lambda_0 \sum_i(1-\gamma_{ij}), $ where $\lambda_{ij}^*$ are set according to the Wardrop equilibrium. For all possible action profiles, the equilibrium rates and the corresponding RGF are given as follows:

%Following give equillirbrium arrival rates anda RGF when each CP either plays SN, NS or NN. 

	\begin{enumerate}
		\item For $(SN,SN)$:
                    $\lambda_{11}^*=(m_1-\overline{m}/2+\lambda_0), \lambda_{21}^*=0, \lambda_{12}^*=(m_2-\overline{m}/2+\lambda_0)$ and $\lambda_{22}^*=0$. \\
                    $RGF_1=1+2\lambda_0/\lambda\quad \mbox{and} \quad  RGF_2=1.$
		
                     \item For $(SN,NS)$:
                     $\lambda_{11}^*=(m_1-\overline{m}/2+\lambda_0), \lambda_{21}^*=0, \lambda_{12}^*=0$ and $\lambda_{22}^*=(m_2-\overline{m}/2+\lambda_0)$. \\
                      $RGF_1=1+\lambda_0/\lambda_{11}^* \quad \mbox{and} \quad RGF_2=1+\frac{\lambda_0}{\lambda_{22}^*}.$
		
                     \item For $ (SN,NN)$:
		$\lambda_{11}^*=m_1-1/\alpha+\lambda_0, \lambda_{21}^*=0, \lambda_{12}^*=m_2-1/(\alpha-c_1
)$ and $\lambda_{22}^*=0$, where $\alpha=c_1/2+1/\overline{m}+\sqrt{c_1^2/4+ 1/\overline{m}^2}$
		and can be bounded as $c_1+ 1/\overline{m} \leq \alpha\leq c_1+ 2/\overline{m}$. \\
		$RGF_1=1+\lambda_0/\lambda \quad \mbox{and} \quad RGF_2=1.$
                    \item For $(NS,SN)$:
		$\lambda_{11}^*=0, \lambda_{21}^*=(m_1-\overline{m}/2+\lambda_0), \lambda_{12}^*=(m_2-\overline{m}/2+\lambda_0)$ and $\lambda_{22}^*=0$.\\
                $RGF_1=1+\frac{\lambda_0}{\lambda_{12}^*} \quad \mbox{and} \quad RGF_2=1+\lambda_0/\lambda_{22}^*.$

                     \item For $(NS,NS)$:
		$\lambda_{11}^*=0, \lambda_{21}^*=(m_1-\overline{m}/2+\lambda_0), \lambda_{12}^*=0$ and $\lambda_{22}^*=(m_2-\overline{m}/2+\lambda_0)$. \\
                      $RGF_1=1 \quad \mbox{and} \quad RGF_2=1+2\lambda_0/\lambda.$

                     \item For $(NS,NN)$:
		$\lambda_{11}^*=0, \lambda_{21}^*=m_1-1/\alpha+\lambda_0, \lambda_{12}^*=m_2-1/(\alpha-c_1)$ and $\lambda_{22}^*=0$. \\
                     $RGF_1=1 \quad \mbox{and} \quad RGF_2=1+\lambda_0/\lambda_{21}^*.$

                     \item For $(NN,SN)$:
		$\lambda_{11}^*=m_1-1/(\alpha-c_1), \lambda_{21}^*=0, \lambda_{12}^*=m_2-1/\alpha+\lambda_0$ and $\lambda_{22}^*=0$.\\
               $RGF_1=1+\lambda_0/\lambda \quad \mbox{and} \quad RGF_2=1.$
 
                     \item For $(NN,NS)$:
		$\lambda_{11}^*=m_1-1/(\alpha-c_1), \lambda_{21}^*=0, \lambda_{12}^*=0$ and $\lambda_{22}^*=m_2-1/\alpha+\lambda_0$. \\
		$RGF_1=1 \quad \mbox{and} \quad RGF_2=1+\lambda_0/\lambda_{22}^*.$
           
                     \item For $(NN,NN)$:
		$\lambda_{11}^*=(m_1-\overline{m}/2), \lambda_{21}^*=0, \lambda_{12}^*=(m_2-\overline{m}/2)$ and $\lambda_{22}^*=0$. \\
		 $RGF_1=1 \quad \mbox{and} \quad RGF_2=1.$

	\end{enumerate}

%Following theorem gives characterization of PNE.
%
%\begin{thm}
%	\label{thm:PNE2}
%Let $\rho, c, \beta^i_j=\beta$ $\forall i,j$ be given. There are only three PNE: $(SN,SN)$, $(NN,NN)$, and $(SN,NN)$, and 
%
%\end{thm}

\begin{figure*}[!h]
	\centering
	\subfloat[]{\includegraphics[scale=0.416]{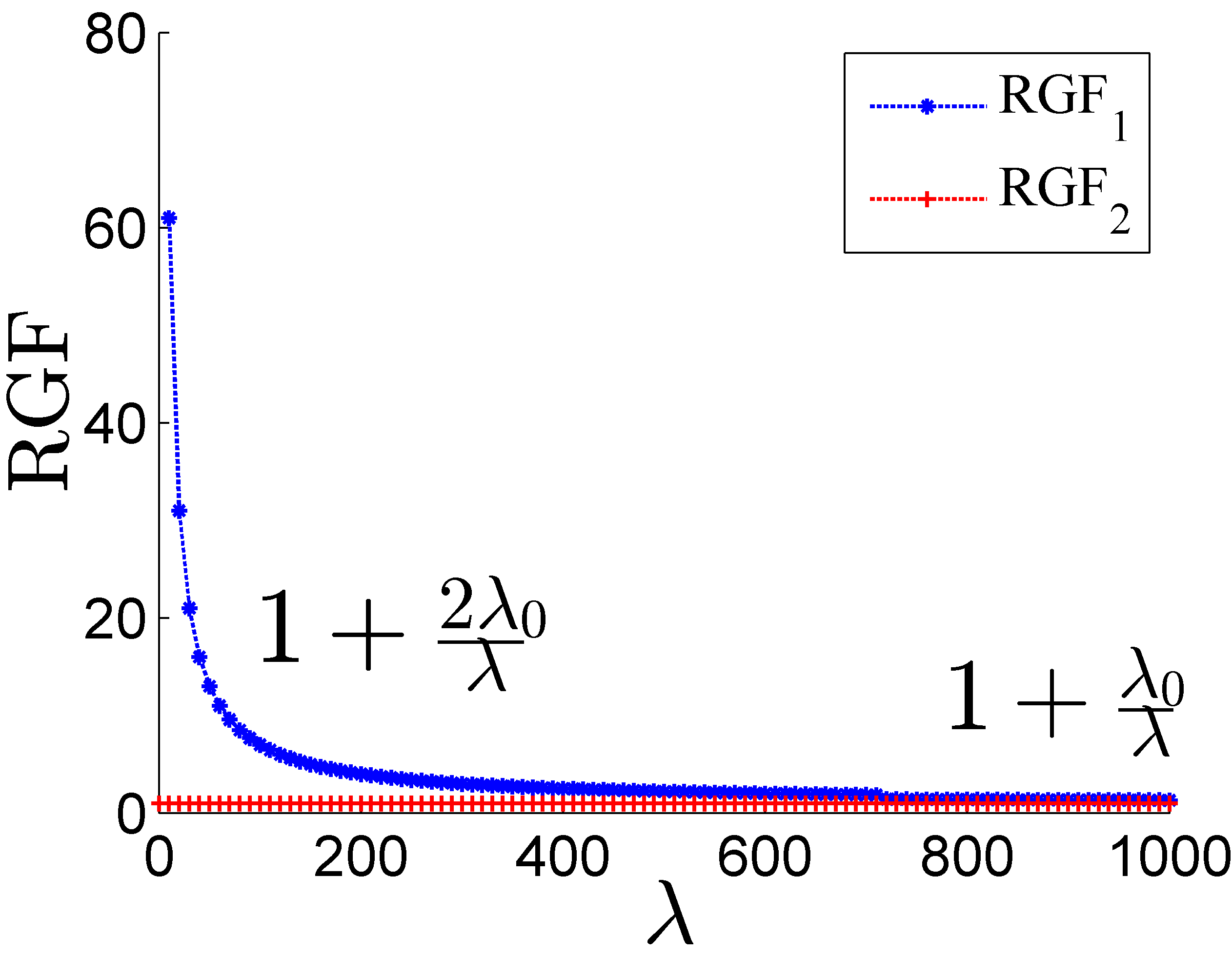}%
		\label{fig:L1}}
	\hspace{0.0000cm}
	\subfloat[]{\includegraphics[scale=0.415]{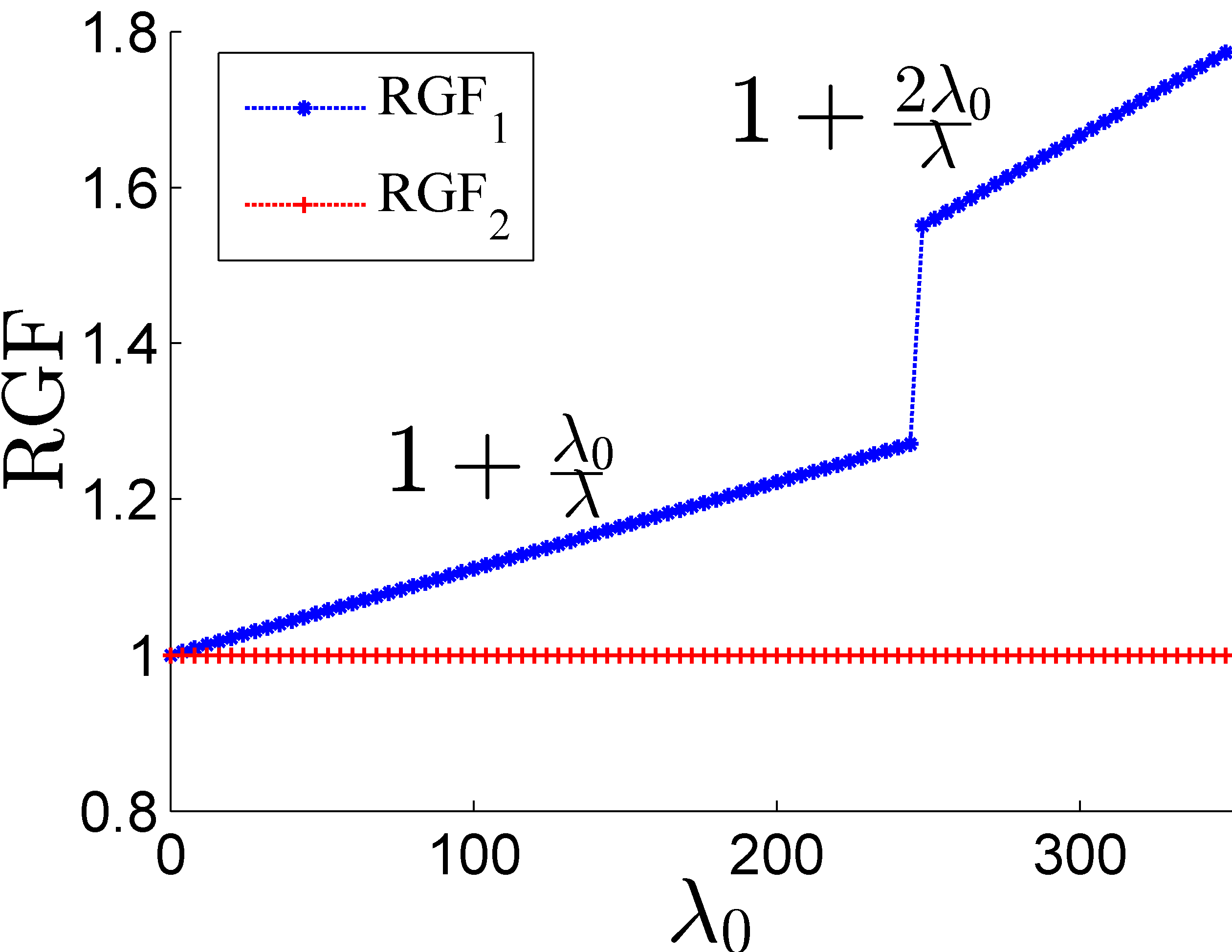}%
		\label{fig:L0}}
	\hspace{0.00cm}
	\subfloat[]{\includegraphics[scale=0.439]{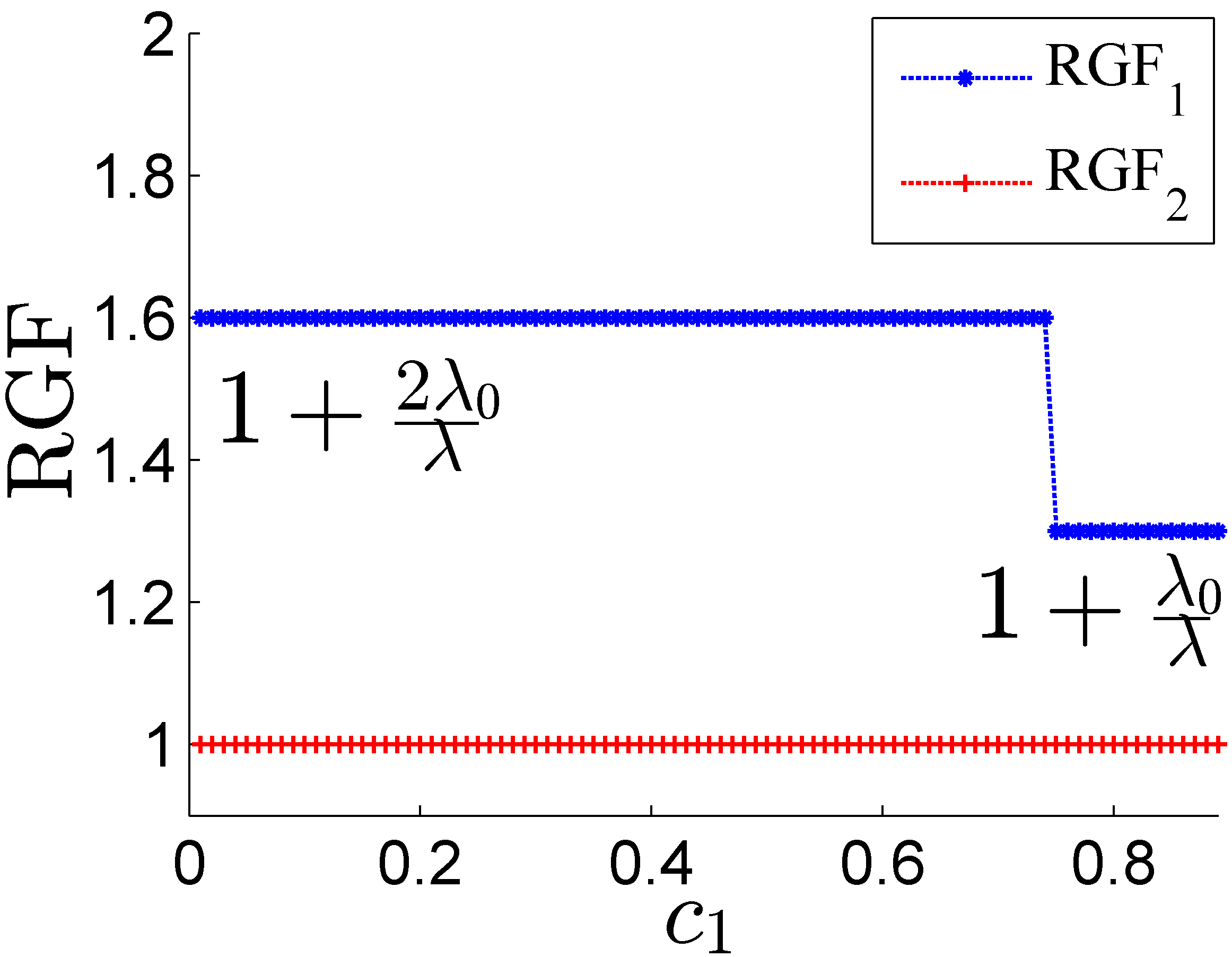}%
		\label{fig:c1}}
	\hspace{0.00cm}
	\subfloat[]{\includegraphics[scale=0.435]{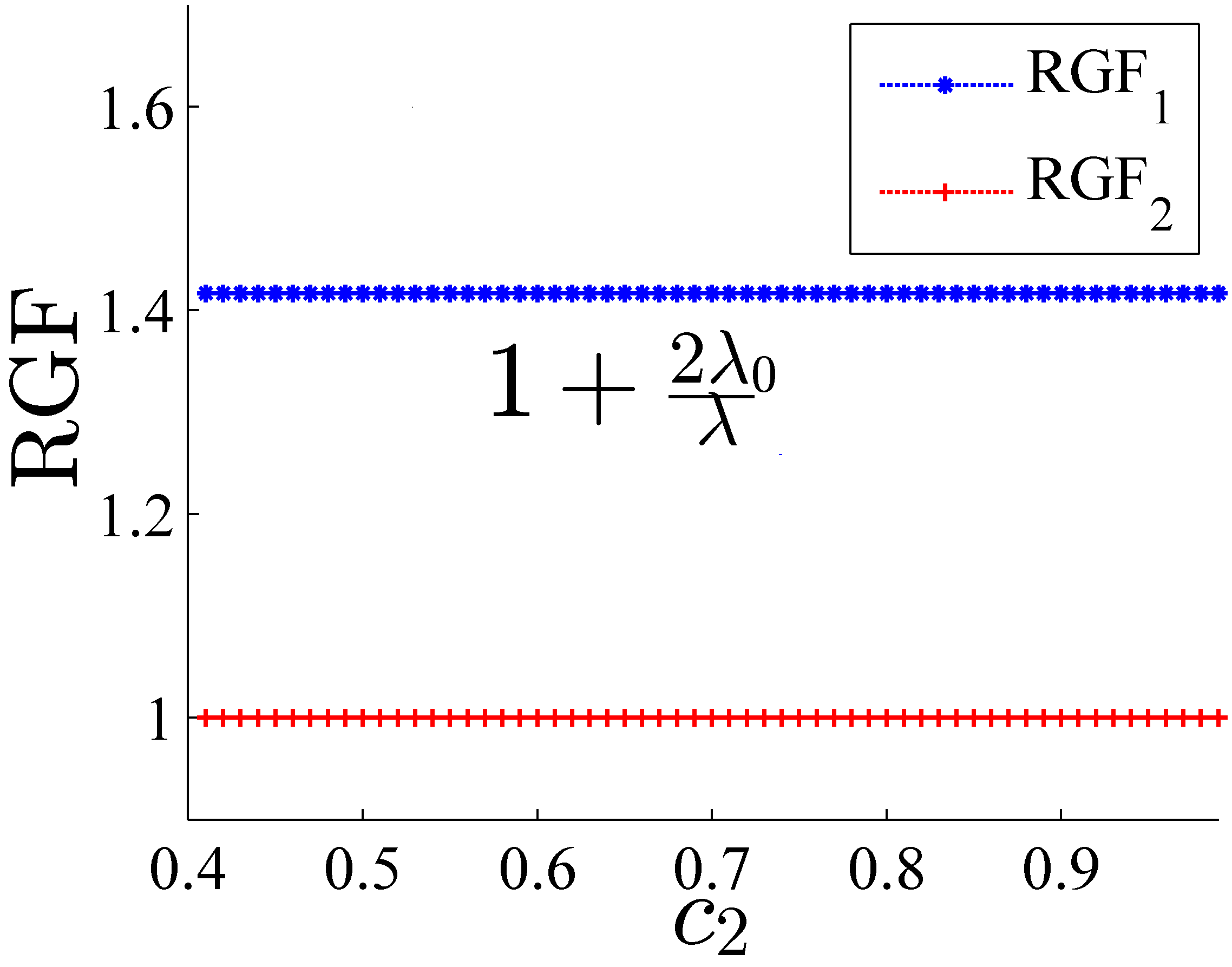}%
		\label{fig:c2}}
	\caption {The change in gain for ISPs and shift in PNE point is being studied with respect to change in parameters $\lambda$, $\lambda_0$, $c_1$ and $c_2$in Figs. \ref{fig:L1}, \ref{fig:L0}, \ref{fig:c1} and \ref{fig:c2} respectively.We set $\beta=1$, $m_1=700$ and $m_2=900$ throughout. In Fig. \ref{fig:L1}, we set $c_1=0.7$, $c_2=0.9$, $\rho=0.9$ and $\lambda_0=300$, in Fig. \ref{fig:L0}, we set $c_1=0.7$, $c_2=0.9$, $\rho=0.9$ and $\lambda=900$, in Fig. \ref{fig:c1}, we set $c_2=0.9$, $\rho=0.8$ $\lambda=1000$ and $\lambda_0=300$ and  in Fig. \ref{fig:c2}, we set $c_1=0.4$, $\rho=0.9$ $\lambda=1200$ and $\lambda_0=250$.}
	\label{fig:RGF2}
\end{figure*}

We illustrate the behavior of RGF for each ISP at different PNE in Figure \ref{fig:RGF2}. Since at PNE, none of the CPs make sponsorship with $ISP_2$ (ISP with higher access price) and the RGF of $ISP_2$ remains the same. Therefore, we focus on the revenue gain of $ISP_1$. It can be seen from Figures \ref{fig:L1}-\ref{fig:c1} that the behavior of the RGF of $ISP_1$ is exactly same as that of single ISP case. And there is no impact of $c_2$ on RGF for both ISPs which is because $c_2>c_1$.

% From Fig. \ref{fig:L1} it is observed that as $\lambda$ increases, RGF decreases and PNE point for CPs get shifted from\{SN,SN\} to \{NN,NS\}. It is inferred that when intrinsic traffic is low both CPs will sponsor to attract more traffic and with higher intrinsic traffic only CP with higher capacity will sponsor. Fig. \ref{fig:L0} shows that RGF initially increases with increase in exogenous traffic ($\lambda_0$) and PNE point get shifted from \{SN,SN\} to \{NN,NS\}.  That is, when endogenous traffic is low both CPs will sponsor to attract more traffic, however, as endogenous traffic increases only CP with higher capacity will sponsor. ISP is also benefitted when higher exogenous traffic is generated. Fig. \ref{fig:c1} shows that as the access price set by $ISP_1$ increases, RGF decreases as PNE point get shifted from \{NN,NS\} to \{NN,NN\}. That is, at higher access price none of the CPs will sponsor and ISP doesn’t get any benefit by setting high access price. While Fig. \ref{fig:c2} shows that as the access price set by $ISP_2$ increases, RGF increases as PNE point get shifted from \{NN,NN\} to \{NN,NS\}. That is, CP will sponsor with ISP with lower access price as the ISP with higher access price increases its cost. This also implies ISP doesn’t get any benefit by setting high access price. That is, there is indirect price war between ISPs and by setting lower access price, they get more benefit. 

%% file: Conclusions.tex
In this work we have analyzed interaction between ISPs, CPs and end users as a complex interacting system where sponsoring of the content (differential pricing) is allowed in the Internet. Our analysis suggests that a CP with poor QoS can attract more end user traffic demand by offering heavy subsidy and earn more revenue than a CP with better QoS if the later do not offer higher subsidy. This implies that differential pricing schemes can be unfair and decentivize the CPs  to improve their QoS through systematic long term investments, but encourage them to focus more on the running costs involving subsidies. Zero-rating schemes thus suit CPs with poor QoS that like to improve their revenues without updating QoS at their facilities.

Our analysis also suggests that overall QoS experienced by end users can worsen in the differential pricing regime -- if a CP with poor QoS offers heavy subsidy on access price, it increases congestion at the CP  level (and reduces at the others) and effectively increases the overall mean delay experienced by the end users in the network. However, if a CP with better QoS offers more subsidy than that of the CPs with lower QoS, the overall delay experienced by the users decreases, making the differential pricing regime favorable to end users.

As our analysis suggests, differential schemes result in unfair distribution of revenues among the CPs and degrade QoS experience for the end users if CPs with poor QoS offer higher subsidy on access price than the CPs with better QoS. Thus, as a policy recommendation, we suggest that the CPs with poor QoS should not be allowed to offer higher subsidy on the access price than that offered by the CPs with higher QoS.  Alternatively, CPs should be only allowed to subsidize the access price in proportion to their QoS guarantees.

Our current model looks at single type of content. As a future work, we will look into a more general model with multiple types of populations corresponding to different content types with different QoS requirement at CPs.

%% file: Appendix.tex
\section*{Proof of Lemma \ref{lma:EquRates}}
	First note that by Assumption (2), $\lambda_i^* >0$ for $i=1,2$. From equation (4), there exists $\alpha>0$ such that
	\[\frac{1}{m_1-\lambda_1^*}+\gamma_1c=\frac{1}{m_2-\lambda_2^*}+\gamma_2 c=\alpha.\]
	Simplifying the above we get:
	\[\lambda_i^* = m_i - \frac{1}{\alpha- \gamma_ic} \mbox{  for } i=1,2.\]
	In order to compute the value of $\alpha$, we use the relation $\lambda_1^*+\lambda_2^*=\lambda$ which yields
	\[m_1 +m_2-\lambda= \frac{1}{\alpha- \gamma_1c}+ \frac{1}{\alpha- \gamma_2c}.\]
	Simplifying the above, we get a quadratic equation in $\alpha$. It is easy to argue that one of the roots is not feasible and the other root gives the required value of $\alpha$. \hfill\IEEEQED %For more details see Appendix.
\section*{Proof of Lemma \ref{lma:DelaySimplify}}
	From (\ref{eqn:EquRates}) for all $i=1,2$ we have
	\[\lambda_i^*=m_i-\frac{1}{\alpha-\gamma_i c}
	\mbox{ and }
	\frac{1}{m_i-\lambda_i^*}=\alpha-c\gamma_i.\]
	Substituting in (\ref{eqn:Delay}) and simplifying we get (\ref{eqn:DealySimpliefied}).
	When $\gamma_i=\gamma$ for all $i$, 
	the minimum cost at equilibrium (from (\ref{eqn:MinCost})) is given as 
	\[\alpha:=\alpha(\gamma, \cdots, \gamma)=c\gamma + \frac{2}{\overline{m}}.\]
	Finally, Substituting the above in (\ref{eqn:DealySimpliefied}) we get (\ref{eqn:DealySymmetric}).%\hfill \IEEEQED

\section*{Proof of Lemma \ref{lma:DelayProp}}
	Substituting value of $\alpha$ from (\ref{eqn:MinCost}) in (\ref{eqn:DealySimpliefied}) and simplifying we have
	\begin{eqnarray*}
		D(c,\gamma_1,\gamma_2)&=&\frac{c(\G_1-\G_2)(m_2-m_1)}{2\lambda} +\frac{m_1+m_2}{\lambda\OL{m}}\\
		&-& \frac{2}{\LA}
		+(m_1+m_2)\frac{\sqrt{(c\overline{m}(\G_1-\G_2))^2+4}}{2\LA \OL{m}}.
	\end{eqnarray*}
	First consider the case $\G_1 \geq \G_2$. It is clear that $D(c,\G_1,\G_2)$ is monotonically increasing in $c$.
	Now consider the case $\G_1 < \G_2$. Differentiating $D(c,\G_1,\G_2)$ with respect to $c$ and simplifying we get
\begin{eqnarray*}
	\frac{\partial D(c,\G_1,\G_2)}{\partial c}&=&\frac{|\G_1-\G_2|(m_1+m_1)}{2\lambda}\times	\\
	&& \hspace{-.5cm} \left \{ -\frac{m_2-m_1}{m_1+m_2}+\frac{c\OL{m}|\G_1-\G_2|}{\sqrt{(c\OL{m}|\G_1-\G_2|^2)+4}} \right\}.
\end{eqnarray*}
It is now easy to verify that the above derivative is positive for all \[c\geq \left( \sqrt{m_2/m_1}-\sqrt{m_1/m_2}\right)/(\OL{m}|\G_1-\G_2|)\] and it is negative otherwise. Hence $D(c,\G_1,\G_2)$ is convex in $c$ with a unique minimum. %\hfill\IEEEQED

\section*{Proof of Theorem \ref{thm:DelayProp}}
From Lemma (\ref{lma:DelayProp}) recall that
\begin{eqnarray*}
	D(c,\gamma_1,\gamma_2)&=&\frac{c(\G_1-\G_2)(m_2-m_1)}{2\lambda} +\frac{m_1+m_2}{\lambda\OL{m}}-\frac{2}{\LA} \\
	&+&(m_1+m_2)\frac{\sqrt{(c\overline{m}(\G_1-\G_2))^2+4}}{2\LA \OL{m}}.
\end{eqnarray*} and
\begin{equation*}
D(c,1,1)= \frac{2}{\OL{m}}.
\end{equation*}	
We have
\begin{eqnarray*}
&&D(c,\gamma_1,\gamma_2) \geq D(c,1,1) \\
&\iff&  \frac{c(\G_1-\G_2)(m_2-m_1)}{2\lambda}  +\frac{m_1+m_2}{\lambda\OL{m}}  \\
&& \quad + (m_1+m_2)\frac{\sqrt{(c\overline{m}(\G_1-\G_2))^2+4}}{2\LA \OL{m}} \geq \frac{2}{\LA} + \frac{2}{\OL{m}}\\
&\iff&(m_1+m_2)\sqrt{(c\overline{m}(\G_1-\G_2))^2+4}  \\ 
&& \quad \geq  2(m_1+m_2)  + c\OL{m}(\G_2-\G_1)(m_2-m_1) \\
&\iff&(c\overline{m}(\G_1-\G_2))^2\geq   4c\OL{m}(\G_2-\G_1)\left(\frac{m_2-m_1}{m_2+m_1}\right) \\ && \quad +(c\OL{m}(\G_2-\G_1))^2\left(\frac{m_2-m_1}{m_2+m_1}\right)^2 \\
&\iff& (c\OL{m}(\G_2-\G_1))^2\left(1-\left(\frac{m_2-m_1}{m_2+m_1}\right)^2\right )  \\
&& \quad \geq  4c\OL{m}(\G_2-\G_1)\left(\frac{m_2-m_1}{m_2+m_1}\right).
\end{eqnarray*}
When $\G_1 > \G_2$, the last inequality holds for all $c$. Hence the first claim is proved. Now consider the case $\G_2> \G_1$. Dividing both sides of the last inequality by $\G_2-\G_1 >0$ and continuing the chain of if and only if conditions, we have
\begin{eqnarray*}
	&&D(c,\gamma_1,\gamma_2) \geq D(c,1,1) \\
	&\iff& c\OL{m}(\G_2-\G_1)\left(1-\left(\frac{m_2-m_1}{m_2+m_1}\right)^2\right )  \geq \;\; 4\left(\frac{m_2-m_1}{m_2+m_1}\right)\\
	&\iff& c \geq \left (\frac{m_2}{m_1}-\frac{m_1}{m_2}\right ) \frac{1}{\OL{m}(\G_2-\G_1)}.%\hfill%\IEEEQED
\end{eqnarray*}

\section*{Proof of Proposition \ref{prop:EquRateProperty}}
	Consider the case $\G_1<\G_2$.
	From (\ref{eqn:MinCost}) we have
	\[\A - c\G_1=\frac{c(\G_2-\G_1)}{2}+\frac{2}{\OL{m}}+\frac{\sqrt{(c\OL{m}(\G_1-\G_2))^2+4}}{2\OL{m}}.\]
	Differentiating both with respect to parameter $c$ we get:
	\[\frac{\partial(\A - c\G_1)}{\partial c}=\frac{(\G_2-\G_1)}{2}+\frac{c\OL{m}(\G_1-\G_2)^2}{2\sqrt{(c\OL{m}(\G_1-\G_2))^2+4}}.\]
	%and
	%\[\frac{\partial(\A - c\G_2)}{\partial c}=\frac{(\G_1-\G_2)}{2}+\frac{c\OL{m}(\G_1-\G_2)^2}{2\sqrt{(c\OL{m}(\G_1-\G_2))^2+4}}.\]
	It is clear that $\frac{\partial(\A - c\G_1)}{\partial c}> 0$ for all $c$. %Rewrite the second derivate above as 
	%\[\frac{\partial(\A - c\G_2)}{\partial c}=\frac{|\G_1-\G_2|}{2}\times \left \{-1 +\frac{c\OL{m}|\G_1-\G_2|}{\sqrt{(c\OL{m}|\G_1-\G_2|)^2+4}}\right \}.\]
	%Now it is clear that $\frac{\partial(\A - c\G_2)}{\partial c}\leq 0$ for all $c$.
	Hence $\alpha -\G_1c$ is monotonically increasing in $c$. The claim follows from (\ref{eqn:EquRates}) and noting that $\lambda_2^*=\lambda-\lambda_1^*$.
	
	\noindent
	Now consider the case $\G_2<\G_1$. From (\ref{eqn:MinCost}) we have
	\[\A - c\G_2=\frac{c(\G_1-\G_2)}{2}+\frac{2}{\OL{m}}+\frac{\sqrt{(c\OL{m}(\G_1-\G_2))^2+4}}{2\OL{m}}.\]
	Following similar steps above we observe that $\frac{\partial(\A - c\G_2)}{\partial c}> 0$ for all $c$. Hence $\alpha -\G_2c$ is monotonically increasing in $c$. The claim follows from (\ref{eqn:EquRates}) and noting that $\lambda_1^*=\lambda-\lambda_2^*$.

\section*{Proof of Proposition \ref{prop:CP1larger}}
	We have
	\[\frac{U_1(\G_1,
		\G_2)}{U_2(\G_2,\G_1)}=\frac{(1-(1-\G_1)\overline{\beta})\lambda_1^*}{(1-(1-\G_2)\overline{\beta})\lambda_2^*}\]
	\begin{eqnarray*}
		\lefteqn{U_1(\G_1,
		\G_2)\geq U_2(\G_2,\G_1)}\\
		&\iff&	\frac{\lambda_2^*}{\lambda_1^*}\leq \frac{(1-(1-\G_1)\overline{\beta})}{(1-(1-\G_2)\overline{\beta})} \\
		&\iff&	\frac{\lambda_2^*+\lambda_1^*}{\lambda_1^*}\leq \frac{(1-(1-\G_1)\overline{\beta})}{(1-(1-\G_2)\overline{\beta})}+1\\
		&\iff& \lambda_1^* \geq \lambda\frac{1}{\frac{1-(1-\G_1)\overline{\beta}}{1-(1-\G_2)\overline{\beta}}+1}.
	\end{eqnarray*}
	%\hfill\IEEEQED
\section*{Proof of Theorem \ref{thm:PNE}}
$(S,S)$ is a PNE: From Table \ref{tab:Utilities}, $(S,S)$ is PNF iff the following two conditions hold
 \begin{eqnarray*}
  (\beta-\rho c)(m_1 - 1/\alpha_{00})\geq \beta (m_1 - 1/(\alpha_{10}-c) \\
	(\beta-\rho c)(m_2 - 1/\alpha_{00})\geq \beta (m_2-2/(\alpha_{01}-c))
	\end{eqnarray*}
	Simplifying these two conditions and using our conventions that $m_1 < m_2$, we get (\ref{eqn:SS}).

	\noindent
	$(N,N)$ is a PNE: From Table \ref{tab:Utilities}, $(N,N)$ is PNE iff the following conditions holds
	\begin{eqnarray*}
\beta(m_1 - (1/\alpha_{11}-c))\geq (\beta-\rho c) (m_1-1/\alpha_{01})\\
 \beta(m_2 - 1/(\alpha_{11}-c))\geq (\beta-\rho c) (m_2 - 1/\alpha_{10}).
		\end{eqnarray*}
Simplifying and using our conventions that $m_1< m_2$, we get (\ref{eqn:NN}).\\
	
	\noindent
	$(S,N)$ is a PNE: From Table \ref{tab:Utilities}, $(S,N)$ is PNF iff the following two conditions hold
	\begin{eqnarray*}
(\beta-\rho c)(m_1 - 1/\alpha_{01})\geq \beta (m_1 - 1/(\alpha_{11}-c)) \\
 (\beta-\rho c)(m_2 - 1/(\alpha_{01}-c))\geq (\beta -\rho c) (m_2-2/\alpha_{00}).
		\end{eqnarray*}
Simplifying these two conditions we get (\ref{eqn:SN}).\\
	
	\noindent
	$(N,S)$ is a PNE: From Table \ref{tab:Utilities}, $(S,N)$ is PNF iff the following two conditions hold
	\begin{eqnarray*}
	\beta (m_1 - 1/(\alpha_{10}-c))\geq (\beta-\rho c)(m_1 - 1/\alpha_{00})\\
	(\beta-\rho c)(m_2 - 1/\alpha_{10})\geq \beta  (m_2-1/(\alpha_{11}-c)).
\end{eqnarray*}
i.e., 
	\begin{eqnarray}
\label{eqn:NS}
C:=\frac{ 1/(\alpha_{10}-c)+\lambda_0-1/\alpha_{00}}{c(m_1+\lambda_0-1/\alpha_{00})} \leq \rho/\beta  \nonumber\\
\rho/\beta  \leq \frac{1/(\alpha_{11}-c)+\lambda_0-1/\alpha_{01}}{c(m_2+\lambda_0-1/\alpha_{01})}=:D.
\end{eqnarray}
We will next argue that $C\leq D$ leads to a contradiction.

We have

\begin{eqnarray*}
\lefteqn{C\leq D} \\
&\iff &\frac{ 1/(\alpha_{10}-c)-1/\alpha_{00}}{c(m_1-1/\alpha_{00})} \leq	\frac{1/(\alpha_{11}-c)-1/\alpha_{01}}{c(m_2-1/\alpha_{01})}\\
&\iff&\left(m_2-\frac{1}{\alpha_{11}-c}\right)\left(m_1-\frac{1}{\alpha_{00}}\right)\\
   && \leq  \left(m_1-\frac{1}{\alpha_{10}-c}\right) \left(m_2-\frac{1}{\alpha_{10}} \right) \\
 &\iff&\left(m_2-\OL{m}/2\right)\left(m_1-\OL{m}/2\right) \\
 && \leq \left(m_1-\frac{1}{\alpha_{10}-c}\right) \left(m_2-\frac{1}{\alpha_{10}} \right) (\mbox{   from Thm. \ref{thm:EquRatesSpecialCases}})\\
 &\iff& m_1\left(\frac{1}{\alpha_{10}}-\OL{m}/2\right)+ m_2\left(\frac{1}{\alpha_{10}-c}-\OL{m}/2\right)\\
 && \leq  \frac{1}{\alpha_{10}(\alpha_{10}-c)}-\frac{\OL{m}^2}{4} (\mbox{  after rearranging}) \\
 &\iff& m_1\left(\frac{1}{\alpha_{10}}-\OL{m}/2\right)+ m_2\left(\frac{1}{\alpha_{10}-c}-\OL{m}/2\right)\\
 && \leq  \frac{\OL{m}^2}{2+\sqrt{(c\OL{m})^2+4}}-\frac{\OL{m}^2}{4}
\end{eqnarray*}
where the last inequality follows by substituting value of $\alpha_{10}$ on RHS and simplifying.
It is clear that $\frac{\OL{m}^2}{2+\sqrt{(c\OL{m})^2+4}}-\frac{\OL{m}^2}{4} <0$. We will next show that $ m_1\left(\frac{1}{\alpha_{10}}-\OL{m}/2\right)+ m_2\left(\frac{1}{\alpha_{10}-c}-\OL{m}/2\right)$ is nonnegative.

From Corollary \ref{corol:MinCostRelations} we have $\alpha_{10}-c\leq \alpha_{00}=2/\OL{m}$. Hence we get 
\begin{eqnarray*}
\lefteqn{m_1\left(\frac{1}{\alpha_{10}}-\OL{m}/2\right)+ m_2\left(\frac{1}{\alpha_{10}-c}-\OL{m}/2\right)} \\
&\geq& m_1\left(\frac{1}{\alpha_{10}}-\OL{m}/2\right)+ m_1\left(\frac{1}{\alpha_{10}-c}-\OL{m}/2\right)\\
&=& m_1 \frac{2\alpha_{10}-c}{\alpha_{10}(\alpha_{10}-c)}-m_1\OL{m}   \mbox{  (after simplifying)}\\
&=&m_1\OL{m}-m_1\OL{m}=0,
\end{eqnarray*}
where the last equality follows after substituting the value of $\alpha_{10}$ and simplifying.
\\

\section*{Proof of Lemma 5}
\noindent
\begin{proof}
Assume $\lambda_{1j}^*>0$ and $\lambda_{2j}^*=0$, then by Wardrop condition we have 
\begin{eqnarray*}
\frac{1}{m_j-\lambda_{j}^*}+\gamma_{1j}c_1 <  \frac{1}{m_j-\lambda_{j}^*}+\gamma_{2j}c_2 
\end{eqnarray*}
where  $\lambda_{j}^*= \lambda_{1j}^*+\lambda_{2j}^*$. Hence $\gamma_{1j}c_1 < \gamma_{2j}c_{2}$.  The other direction follows by noting that $\lambda_j^* \neq 0 \forall j$ and applying the Wardrop conditions.  Proof of the other items is similar.  
%\begin{eqnarray*}
%\frac{1}{m_j-\lambda_{j}^*}+\gamma_{2j}c_2 \leq \min_{i,j=1,2} \left \{ \frac{1}{m_j-\lambda_{j}^*}+\gamma_{ij}c_i \right \}\\
%\end{eqnarray*}
%\end{proof}
%where  $\lambda_{i}^*= \lambda_{1i}^*+\lambda_{2i}^*$. 
\end{proof}

\section*{Proof of Theorem \ref{thm:PNE2}}
\noindent
 $(SN,SN)$ is PNE iff the following four conditions hold
 \begin{eqnarray*}
(\beta-\rho c_1)(m_1-\overline{m}/2+\lambda_0) \geq(\beta-\rho c_2)(m_1-\overline{m}/2+\lambda_0)\\
(\beta-\rho c_1)(m_1-\overline{m}/2+\lambda_0) \geq \beta(m_1-1/(\alpha-c_1)\\
(\beta-\rho c_1)(m_2-\overline{m}/2+\lambda_0)\geq (\beta-\rho c_2)(m_2-\overline{m}/2+\lambda_0)\\
(\beta-\rho c_1)(m_2-\overline{m}/2+\lambda_0)\geq \beta(m_2-1/(\alpha-c_1)
	\end{eqnarray*}
Solving first two inequalities, we get 
 \begin{eqnarray*}
c_2 \geq c_1; \rho/\beta \leq \frac{1/(\alpha-c_1)-\overline{m}/2+\lambda_0}{c_1(m_1-\overline{m}/2+\lambda_0)}
	\end{eqnarray*}

Solving second two inequalities, we get 
 \begin{eqnarray*}
c_2 \geq c_1 ; \rho/\beta \leq \frac{1/(\alpha-c_1)-\overline{m}/2+\lambda_0}{c_1(m_2-\overline{m}/2+\lambda_0)}
\end{eqnarray*}
	Using our conventions that $m_1 \leq m_2$ , we get (\ref{eqn:SNSN}).

	\noindent
	$(NN,NN)$ is PNE iff the following four conditions holds
	\begin{eqnarray*}
\beta(m_1-\overline{m}/2))\geq(\beta-\rho c_1)(m_1-1/\alpha+\lambda_0)\\
\beta(m_1-\overline{m}/2)\geq (\beta-\rho c_2)(m_1-1/\alpha+\lambda_0)\\
\beta(m_2-\overline{m}/2) \geq (\beta-\rho c_2)(m_2-1/\alpha+\lambda_0)\\
\beta(m_2-\overline{m}/2) \geq (\beta-\rho c_1)(m_2-1/\alpha+\lambda_0)
		\end{eqnarray*}
Solving first two inequalities, we get 
 \begin{eqnarray*}
\rho/\beta \geq \frac{\overline{m}/2-1/\alpha+\lambda_0}{c_1(m_1-1/\alpha+\lambda_0)}\\
\rho/\beta \geq \frac{\overline{m}/2-1/\alpha+\lambda_0}{c_2(m_1-1/\alpha+\lambda_0)}
	\end{eqnarray*}

Solving second two inequalities, we get 
 \begin{eqnarray*}
\rho/\beta \geq \frac{\overline{m}/2-1/\alpha+\lambda_0}{c_2(m_2-1/\alpha+\lambda_0)}\\
\rho/\beta \geq \frac{\overline{m}/2-1/\alpha+\lambda_0}{c_1(m_2-1/\alpha+\lambda_0)}
\end{eqnarray*}

Using our conventions that $m_1 \leq m_2$ ; $c_1 \leq c_2$, we get (\ref{eqn:NNNN}).\\

$(SN,NN)$ is PNE iff following conditions hold\\
\begin{eqnarray*}
(\beta-\rho c_1)(m_1-1/\alpha+\lambda_0) \geq (\beta-\rho c_2)(m_1-1/\alpha+\lambda_0) \\
(\beta-\rho c_1)(m_1-1/\alpha+\lambda_0) \geq \beta(m_2-\overline{m}/2) \\
\beta(m_2-1/(\alpha-c_1)) \geq (\beta-\rho c_1)(m_2-\overline{m}/2+\lambda_0)\\
\beta(m_2-1/(\alpha-c_1)) \geq (\beta-\rho c_2)(m_2-\overline{m}/2+\lambda_0)
		\end{eqnarray*}
Solving first two inequalities, we get 
 \begin{eqnarray*}
c_2 \geq c_1 ; \rho/\beta \leq \frac{1/\alpha-\overline{m}/2+\lambda_0}{c_1(m_1-\overline{m}/2+\lambda_0)}
	\end{eqnarray*}

Solving second two inequalities, we get 
 \begin{eqnarray*}
\rho/\beta \geq \frac{1/(\alpha-c_1)-\overline{m}/2+\lambda_0}{c_1(m_2-\overline{m}/2+\lambda_0)}\\
\rho/\beta \geq \frac{1/(\alpha-c_1)-\overline{m}/2+\lambda_0}{c_2(m_2-\overline{m}/2+\lambda_0)}
\end{eqnarray*}
	Using our conventions that $m_1 \leq m_2$ and $c_1 \leq c_2$, we get (4).
 %\begin{eqnarray*}
%\frac{1/(\alpha-c_1)-\overline{m}/2}{c_1(m_2-\overline{m}/2)} \leq \rho/\beta \leq \frac{1/\alpha-\overline{m}/2}{c_1(m_1-%\overline{m}/2)}
%\end{eqnarray*}

	\noindent
$(NN,SN)$ is PNE iff the following four conditions holds
\begin{eqnarray*}
\beta(m_1-1/(\alpha-c_1)) \geq (\beta-\rho c_2)(m_1-\overline{m}/2+\lambda_0)\\
\beta(m_1-1/(\alpha-c_1)) \geq (\beta-\rho c_1)(m_1-\overline{m}/2+\lambda_0)\\
(\beta-\rho c_1)(m_2-1/\alpha+\lambda_0) \geq (\beta-\rho c_2)(m_2-1/\alpha)\\
(\beta-\rho c_1)(m_2-1/\alpha+\lambda_0) \geq \beta(m_2-\overline{m}/2)
		\end{eqnarray*}
Solving first two inequalities, we get 
 \begin{eqnarray*}
\rho/\beta \geq \frac{1/(\alpha-c_1)-\overline{m}/2+\lambda_0}{c_2(m_1-\overline{m}/2+\lambda_0)}\\
\rho/\beta \geq \frac{1/(\alpha-c_1)-\overline{m}/2+\lambda_0}{c_1(m_1-\overline{m}/2+\lambda_0)}:=A1
	\end{eqnarray*}
Solving second two inequalities, we get 
 \begin{eqnarray*}
c_1\leq c_2; \rho/\beta \leq \frac{\overline{m}/2-1/\alpha+\lambda_0}{c_1(m_2-1/\alpha+\lambda_0)}:=B1
\end{eqnarray*}
Using our conventions that $m_1 \leq m_2$ and $c_1 \leq c_2$, we get 
 \begin{eqnarray*}
\frac{1/(\alpha-c_1)-\overline{m}/2+\lambda_0}{c_1(m_1-\overline{m}/2+\lambda_0)} \leq \rho/\beta \leq \frac{\overline{m}/2-1/\alpha+\lambda_0}{c_1(m_2-1/\alpha+\lambda_0)}
\end{eqnarray*}
which is not true because $A1>B1$ (the proof is exactly on same lines to that of $C>D$).\\	
Other strategies cannot be PNE. This is because the conditions to be PNE gives $c_1\geq c_2$ which is contradiction to our assumption.

%	\begin{eqnarray*}
%%\begin{eqnarray*}
%%\lefteqn{A\leq B} \\
%%&\iff &\frac{ 1/(\alpha_{10}-c)-1/\alpha_{00}}{c(m_2-1/\alpha_{00})} \leq	\frac{1/(\alpha_{11}-c)-1/\alpha_{01}}{c(m_1-1/\alpha_{01})}\\
%%&\iff&\left(m_1-\frac{1}{\alpha_{11}-c}\right)\left(m_2-\frac{1}{\alpha_{00}}\right)\\
% %  && \leq  \left(m_2-\frac{1}{\alpha_{10}-c}\right) \left(m_1-\frac{1}{\alpha_{10}} \right) \\
% %&\iff&\left(m_2-\overline{m}/2\right)\left(m_1-\overline{m}/2\right) \\
% %& \leq \left(m_2-\frac{1}{\alpha_{10}-c}\right) \left(m_1-\frac{1}{\alpha_{10}} \right) (\mbox{   from Thm. \ref{thm:EquRatesSpecialCases}})\\
% %&\iff& m_2\left(\frac{1}{\alpha_{10}}-\overline{m}/2\right)+ m_1\left(\frac{1}{\alpha_{10}-c}-\overline{m}/2\right)\\
% %&& \leq  \frac{1}{\alpha_{10}(\alpha_{10}-c)}-\frac{\overline{m}^2}{4} (\mbox{  after rearranging}) \\
% %&\iff& m_2\left(\frac{1}{\alpha_{10}}-\overline{m}/2\right)+ m_1\left(\frac{1}{\alpha_{10}-c}-\overline{m}/2\right)\\
% %&& \leq  \frac{\overline{m}^2}{2+\sqrt{(c\overline{m})^2+4}}-\frac{\overline{m}^2}{4}\\
%  %&\iff& \frac{\lambda^2-(m_1+m_2)^2}{4}\\
% %&& \leq  \frac{1-m_1\alpha_{10}-m_2(\alpha_{10}-c)}{\alpha_{10}(\alpha_{10}-c)}
%%\end{eqnarray*}